# Nonlinear Electrostatics. The Poisson-Boltzmann Equation

**C. G. Gray* and P. J. Stiles#**

***Department of Physics, University of Guelph, Guelph, ON N1G2W1, Canada***

*(cgray@uogueph.ca)*

**#*Department of Molecular Sciences, Macquarie University, NSW 2109, Australia***

*(peter.stiles@mq.edu.au)*

The description of a conducting medium in thermal equilibrium, such as an electrolyte solution or a plasma, involves nonlinear electrostatics, a subject rarely discussed in the standard electricity and magnetism textbooks. We consider in detail the case of the electrostatic double layer formed by an electrolyte solution near a uniformly charged wall, and we use mean-field or Poisson-Boltzmann (PB) theory to calculate the mean electrostatic potential and the mean ion concentrations, as functions of distance from the wall. PB theory is developed from the Gibbs variational principle for thermal equilibrium of minimizing the system free energy. We clarify the key issue of *which* free energy (Helmholtz, Gibbs, grand, …) should be used in the Gibbs principle; this turns out to depend not only on the specified conditions in the bulk electrolyte solution (e.g., fixed volume or fixed pressure), but also on the specified surface conditions, such as fixed surface charge or fixed surface potential. Despite its nonlinearity the PB equation for the mean electrostatic potential can be solved analytically for planar or wall geometry, and we present analytic solutions for both a full electrolyte, and for an ionic solution which contains only counterions, i.e. ions of sign opposite to that of the wall charge. This latter case has some novel features. We also use the free energy to discuss the inter-wall forces which arise when the two parallel charged walls are sufficiently close to permit their double layers to overlap. We consider situations where the two walls carry equal charges, and where they carry equal and opposite charges.

## INTRODUCTION

Most electricity and magnetism textbooks, and pedagogical journal articles, discuss only linear electrostatics. That is, the charge density of the medium at a point $\boldsymbol{x}$ is either assumed to be fixed, or to be perturbed linearly by the electric field $\boldsymbol{E}(\boldsymbol{x})$. The subject of nonlinear dielectric media has a large literature which is mostly confined to specialist journals and monographs[1-3]. Another important case involving nonlinear electrostatics is a conducting medium in thermal equilibrium such as an electrolyte solution or a plasma. For example, a solution of sodium chloride dissolved in water produces hydrated ions $\mathrm{Na}^+$ and $\mathrm{Cl}^-$. In the absence of an applied field, the solution is spatially uniform and overall neutral but locally the $\mathrm{Cl}^-$ ions tend to cluster around the $\mathrm{Na}^+$ ions, and vice-versa, resulting in a rapid screening out of the electric field around any given ion. The screening occurs essentially over a characteristic distance $\ell_\mathrm{D}$, called the Debye length, where

$$\ell_\mathrm{D} = \left( \frac{\varepsilon k_\mathrm{B} T}{2 e^2 \bar{c}} \right)^{\frac{1}{2}} \quad . \tag{1}$$

Here $\varepsilon = \varepsilon_\mathrm{r}\, \varepsilon_0$ is the permittivity of water (we use SI units), $\varepsilon_\mathrm{r}$ the dimensionless dielectric constant ($\varepsilon_\mathrm{r}$ is about 80 for water at room temperature), $\varepsilon_0$ the permittivity of free space, $e$ is the magnitude of the electron charge, $k_\mathrm{B}$ is Boltzmann's constant, $T$ the temperature, and $\bar{c}$ the bulk concentration or number density (in ions per $\mathrm{m}^3$) of the $\mathrm{Na}^+$ ions (or the $\mathrm{Cl}^-$ ions). Thus, each ion is, on average, surrounded by a spherically symmetric screening cloud of radius of order $\ell_\mathrm{D}$, which is about 1 nm at room temperature for a bulk ion concentration of 100 mM. As we discuss later, we have here assumed the so-called primitive or implicit solvent model with the solvent (assumed to be water in this example) taken to be a dielectric continuum with permittivity $\varepsilon$, and we have also assumed point ions. A key assumption is that mean field theory applies, i.e., we have assumed that the electrostatic field at a position $\boldsymbol{x}$ from a central ion is the Coulomb field of the central ion together with the thermal average field due to all the other ions. Any ion near the central one is assumed to sense this total mean field. The ion-ion correlations in a screening cloud due to Coulombic interactions are neglected. Eq (1) was derived by Debye and Hückel[4] based on the assumptions just stated. One writes down two equations: first the rigorous Poisson equation relating the total mean potential $\phi(\boldsymbol{x})$ at a position $\boldsymbol{x}$ from a central ion to the total mean charge density $\rho(\boldsymbol{x})$ due to the central ion (here called "fixed" for reasons clarified in the next paragraph), and to the other (here called "mobile") ions,

$$\nabla^2\phi(\boldsymbol{x}) = -\rho(\boldsymbol{x})/\varepsilon \quad , \tag{2}$$

with total mean charge density

$$\rho(\boldsymbol{x}) = \rho_{\mathrm{f}}(\boldsymbol{x}) + \rho_{\mathrm{m}}(\boldsymbol{x}) \quad , \tag{3}$$

where $\rho_{\mathrm{f}}(\boldsymbol{x})$ is the charge density of the fixed ion (a delta function, e.g., $e\,\delta(\boldsymbol{x})$ for a central positive ion (cation)) and $\rho_{\mathrm{m}}(\boldsymbol{x})$ is the local mean charge density of the mobile ions. By definition, we have

$$\rho_{\mathrm{m}}(\boldsymbol{x}) = e\,c_{+}(\boldsymbol{x}) - e\,c_{-}(\boldsymbol{x}) \quad , \tag{4a}$$

where $c_{\pm}(\boldsymbol{x})$ are the local mean concentrations of the mobile ions, and, assuming Boltzmann relations in the approximate form $c_{\pm}(\boldsymbol{x}) = \bar{c}\,\exp\left(\mp\beta\,e\,\phi(\boldsymbol{x})\right)$ we thus have,

$$\rho_{\mathrm{m}}(\boldsymbol{x}) = e\,\bar{c}\,\exp\left(-\beta\,e\,\phi(\boldsymbol{x})\right) - e\,\bar{c}\,\exp\left(+\beta\,e\,\phi(\boldsymbol{x})\right) \quad , \tag{4b}$$

where $\beta = 1/k_{\mathrm{B}}T$ and $\bar{c}$ is the bulk concentration of the electrolyte at large distances from the central ion where $\phi(\boldsymbol{x})$ is assumed to approach zero. From (2) we see that $\phi$ depends on $\rho$, and from (4b) $\rho$ depends on $\phi$, so that we have two equations that must be solved self-consistently for these two quantities. Eqs (2-4) can be combined into a single nonlinear partial differential equation for $\phi(\boldsymbol{x})$ , called the Poisson-Boltzmann (PB) equation:

$$\nabla^2\phi(\boldsymbol{x}) = (e\bar{c}/\varepsilon)\Big(\exp\big(\beta e\phi(\boldsymbol{x})\big) - \exp\big(-\beta e\phi(\boldsymbol{x})\big)\Big) - (1/\varepsilon)\rho_{\mathrm{f}}(\boldsymbol{x}) \quad . \tag{5}$$

For weak electric fields one can linearize (5) in $\phi(\boldsymbol{x})$, thereby deriving the linearized Poisson-Boltzmann (LPB) equation,

$$\nabla^2\phi(\boldsymbol{x}) = \kappa^2\phi(\boldsymbol{x}) - (1/\varepsilon)\rho_{\mathrm{f}}(\boldsymbol{x}) \ , \tag{6}$$

where $\kappa = 1/\ell_{\mathrm{D}}$ , as first done by Debye and Hückel[4] and thus also called the DH equation. Because of the spherical symmetry of simple ions, the various functions of $\boldsymbol{x}$ above are in fact functions of just the radial coordinate $r = |\boldsymbol{x}|$, and (5) and (6) reduce to ordinary differential equations with the radial part of the Laplacian, $r^{-1}d^2/dr^2\, r = d^2/dr^2 + (2/r)d/dr$, replacing $\nabla^2$. We keep the general forms, however, for reasons that will become clear in the next paragraph. For $\rho_{\mathrm{f}}(\boldsymbol{x}) = e\delta(\boldsymbol{x})$ the appropriate solution of (6) is easily verified to be $\phi(r) = \left(e/4\pi\varepsilon r\right)\exp(-\kappa r)$ , variously called the DH, screened Coulomb or Yukawa potential, and which shows explicitly that $\ell_{\mathrm{D}}$ is the screening length, at least for weak fields. It is also the Green function for the operator $\left(\nabla^2 - \kappa^2\right)$ occurring in equation (6).

Interestingly, the full nonlinear PB equation (5) was derived before Debye and Hückel by Gouy[5] and Chapman[6] who considered the problem of a diffuse double layer formed by an electrolyte solution in the presence of a charged wall, e.g., an electrode or a charged biomembrane. The only formal change is that in this case the fixed charge density $\rho_{\mathrm{f}}(\boldsymbol{x})$ is now that due to the charged wall, taken in the simplest case to be a wall with uniform surface charge density $\sigma$, and $\boldsymbol{x}$ now represents an arbitrary point in the fluid. In this system the screening cloud due to the mobile ions forms a diffuse layer with net charge equal and opposite to that of the wall charge, thus producing the double layer, of infinite extent but decaying essentially exponentially as a function of distance from the wall with effective width of the order of $\ell_{\mathrm{D}}$ for not too large $\sigma$. (We assume here an infinite system with one charged wall. Later we also consider finite systems between two charged walls.) We will focus on (5) in the present paper for this double layer. In this case again because of system symmetry $\phi(\boldsymbol{x})$ and $\rho(\boldsymbol{x})$ reduce to functions of just $x$ (the coordinate orthogonal to the charged wall), the full Laplacian can be replaced by $d^2/dx^2$, and (5) reduces to an ordinary nonlinear differential equation. As we shall see, despite its nonlinearity it can be solved analytically for this Gouy-Chapman (wall) geometry. For other geometries, e.g., charged cylindrical or spherical surfaces, numerical methods[7,8] are needed for the full nonlinear PB equation even for the standard simple geometries[9]. Such nonplanar geometries have been used for modelling double layers around cylindrical electrodes, colloids, and charged macromolecules such as proteins and DNA.

As mentioned, standard textbook[10-14] and pedagogical journal[15-22] discussions of the PB (and LPB) equations are uncommon. But because of the importance of electrostatic screening in many systems, there are discussions in a host of monographs for specialized fields including biophysics[9, 23-34], surface science[35-42], chemical physics[43,44], polymer physics[45], plasma physics[46-50], solid state physics[51,52], condensed matter physics[53], many-body theory[54-58], thermodynamics[42,43,59,60], statistical mechanics[61-69], liquid state theory[70-73], electrolyte solutions[74-78], electrochemistry[69,79-85], soft matter[86-91], physical chemistry[92-94], biophysical chemistry[9,95], biochemistry[96], medical physics[97], physiology[98,99], molecular biology[100] , colloids[39,101-105], applied mathematics[8,106,107], materials science[108] and technology[109].

There is also a large literature on theories going beyond primitive model PB[9,61,67,70,71,88,89,91,101,110-112] by including ion-ion correlations, finite ionic size and molecular structure of the solvent (assumed above to be water), nonuniformities of the wall charges, additional intermolecular and wall forces. We do not discuss these improvements here, nor do we discuss the quantum generalization of classical PB theory, but we indicate in the references some quantum discussions[51-58,62]. Thus, for electrons in bulk metals and at metallic surfaces, at not too high temperatures the relevant mean field generalization is the Thomas-Fermi (TF) theory of screening, where the screening length $\ell_{\mathrm{TF}}$ is essentially (1) with $k_{\mathrm{B}}T$ replaced by the Fermi energy $E_{\mathrm{F}}$.

We focus on the double layer problem and carry out two main tasks. Firstly, we give a more systematic and physically rigorous derivation of (5) based on the Gibbs variational principle for thermal equilibrium[113,114], i.e., minimizing the system free energy. This has been done before [70,101,107,110,115-126], but we simplify and clarify earlier derivations by carefully identifying *which* free energy (e.g., Helmholtz, Gibbs, grand, …) is to be minimized. It turns out this depends not only on the assumed conditions of the bulk solution (e.g., fixed ion numbers, volume, and temperature), but also on the surface or boundary conditions chosen for the system, e.g., fixed wall charge or fixed wall potential. Secondly, we use free energy to derive the force on one of the walls when two opposing parallel walls of the container, or two fully immersed parallel surfaces, each has a double layer and the walls or surfaces are sufficiently close to permit overlap of their respective double layers. We consider in detail two walls of identical charge[101,127-128] and then discuss more briefly two walls of opposite charge[128-133]. We also briefly discuss the use of free energy to correct a recent PB calculation[134] which erroneously predicted that the double layer collapses to finite width as the surface charge on a wall is raised beyond a threshold value. Clarke and Stiles[134] were led astray in part by using a free energy inappropriate for their assumed boundary condition.

The paper also includes the analytical solution to the PB equation for a single charged wall, both for a full electrolyte solution and for a solution containing only counterions; the latter case has some novel features. In the case of a single charged wall we also use PB theory to derive simple analytical expressions for observables such as the equilibrium values of various free energies and the differential capacitance. For two charged walls the appendices give convenient algorithms for determining the quantities necessary to calculate the interwall-force; the algorithms provide a simpler route to the required inputs to the force calculations than the standard procedure of fully solving the PB equation numerically. Numerical and graphical examples are given to illustrate some of the results for the one- and two-wall problems.

## II. DERIVATION OF THE PB EQUATION FROM THE HELMHOLTZ FREE ENERGY *F*

We consider a 1:1 univalent electrolyte solution with $N_+$ cations and $N_-$ anions, in thermal equilibrium at a temperature *T* in a cubic container of volume $V = L^3$. The left wall at $x = 0$ is assumed to have a fixed uniform positive charge with surface density $\sigma$. This surface density may arise from various sources, e.g. from preferential adsorption of cations from the solution, or from the positive plate of a battery. A diffuse double layer forms at the $x = 0$ wall. In the simplest manifestation of the primitive model all ions in solution are represented as point charges and the solvent is replaced by a continuous dielectric fluid of uniform electric permittivity $\varepsilon$. The Helmholtz free energy *F* of this non-uniform system can be written as a volume integral of a free energy density $f(\boldsymbol{x})$, i.e.,

$$F = \int_V f(\boldsymbol{x})\,dV \quad . \tag{7}$$

The free energy $F = U - TS$ contains energetic and entropic components, where *U* is the internal energy and *S* the entropy. We approximate both components and write

$f(\boldsymbol{x}) = f_{\rm el}(\boldsymbol{x}) + f_{\rm id}(\boldsymbol{x})$, where[135]

$$f_{\rm el}(\boldsymbol{x}) = \tfrac{1}{2}\rho(\boldsymbol{x})\phi(\boldsymbol{x}) \tag{8}$$

is the electrostatic potential energy in the mean-field approximation[136] in which

$$\phi(\boldsymbol{x}) = \frac{1}{4\pi\varepsilon}\int_V \frac{\rho(\boldsymbol{x}')}{|\boldsymbol{x}-\boldsymbol{x}'|}dV' \tag{9}$$

is the mean Coulomb potential at $\boldsymbol{x}$ due to the mean charge density $\rho$ of the mobile electrolyte charges and the fixed wall charges. The ideal term

$$f_{\mathrm{id}}(\boldsymbol{x}) = \beta^{-1}c_+(\boldsymbol{x})\left[\ln\left(c_+(\boldsymbol{x})\Lambda^3\right)-1\right]+\beta^{-1}c_-(\boldsymbol{x})\left[\ln\left(c_-(\boldsymbol{x})\Lambda^3\right)-1\right] \tag{10}$$

contains the thermally averaged kinetic energy and the ideal-gas Sackur-Tetrode entropic terms[137] (which contain no entropic correlation) for the solution cations and anions. In (10) the thermal de Broglie wavelength $\Lambda = h/\left(2\pi m k_{\mathrm{B}}T\right)^{\frac{1}{2}}$ where $h$ is the Planck constant and all ions are assumed to have equal mass $m$.

We now introduce a Helmholtz free energy $F[c_+,c_-]$ as a functional of the cation and anion density fields $c_+(\boldsymbol{x})$ and $c_-(\boldsymbol{x})$ (in ions per $\mathrm{m}^3$): other fields can be expressed in terms of them. In (10) $f_{\mathrm{id}}(\boldsymbol{x})$ is already expressed in terms of $c_+$ and $c_-$. To express $f_{\mathrm{el}}(\boldsymbol{x})$ in terms of these ionic concentrations we combine (8) and (9) and leave implicit that the charge density $\rho$ is expressed in terms of the concentrations by (3) and (4a). Our functional $F[c_+,c_-]$ is thus

$$F[c_+,c_-] = \int_V\int_V \frac{\rho(\boldsymbol{x})\rho(\boldsymbol{x}')}{8\pi\varepsilon\,|\boldsymbol{x}-\boldsymbol{x}'|}dV'\,dV + \beta^{-1}\int_V\left\{c_+(\boldsymbol{x})\left[\ln\left(c_+(\boldsymbol{x})\Lambda^3\right)-1\right]+c_-(\boldsymbol{x})\left[\ln\left(c_-(\boldsymbol{x})\Lambda^3\right)-1\right]\right\}dV \tag{11}$$

By the Gibbs variational principle[113,114], the free energy functional, here expressed as a density or concentration functional, takes its equilibrium value for the values of the ionic densities that minimize it, subject to the constraints of fixed numbers $N_+ = \int_V c_+(\boldsymbol{x})dV$ of cations and $N_- = \int_V c_-(\boldsymbol{x})dV$ of anions in the solution. We use the standard method of Lagrange multipliers to perform this constrained minimization. Thus, we extremize a functional $J[c_+,c_-] = F[c_+,c_-] - \int_V\left(\lambda_+c_+ + \lambda_-c_-\right)dV$ with no constraints where $\lambda_+$ and $\lambda_-$ are constant Lagrange multipliers and $\rho(\boldsymbol{x})$ is expressed in terms of the ionic densities by (3) and (4a). Varying $c_+$ and $c_-$ we thus find for $\delta J[c_+,c_-]$:

$$\delta J = \int_V \int_V \frac{\left(\delta\rho(\boldsymbol{x})\,\rho(\boldsymbol{x}') + \rho(\boldsymbol{x})\delta\rho(\boldsymbol{x}')\right)}{8\pi\varepsilon\,|\boldsymbol{x}-\boldsymbol{x}'|} dV'\,dV$$
$$+\int_V \left\{\delta c_+(\boldsymbol{x})\left[\beta^{-1}\ln\left(c_+(\boldsymbol{x})\Lambda^3\right)-\lambda_+\right]+\delta c_-(\boldsymbol{x})\left[\beta^{-1}\ln\left(c_-(\boldsymbol{x})\Lambda^3\right)-\lambda_-\right]\right\}dV. \quad (12)$$

The two terms in the double integral are equal by symmetry, and in the first term we eliminate $\rho(\boldsymbol{x}')$ by re-introducing $\phi(\boldsymbol{x})$ using (9) to give

$$\delta J = \int_V dV\left\{\delta\rho(\boldsymbol{x})\phi(\boldsymbol{x}) + \delta c_+(\boldsymbol{x})\left[\beta^{-1}\ln\left(c_+(\boldsymbol{x})\,\Lambda^3\right)-\lambda_+\right]+\delta c_-(\boldsymbol{x})\left[\beta^{-1}\ln\left(c_-(\boldsymbol{x})\,\Lambda^3\right)-\lambda_-\right]\right\}\ . \quad (13)$$

From (3) and (4a), we have $\delta\rho(\boldsymbol{x}) = e\left(\delta c_+(\boldsymbol{x})-\delta c_-(\boldsymbol{x})\right)$. As we require $\delta J = 0$ for arbitrary variations $\delta c_+(\boldsymbol{x})$ and $\delta c_-(\boldsymbol{x})$ in both $c_+(\boldsymbol{x})$ and $c_-(\boldsymbol{x})$, these two conditions yield two Euler-Lagrange equations

$$e\phi(\boldsymbol{x}) + \beta^{-1}\ln\left(c_+(\boldsymbol{x})\Lambda^3\right) = \lambda_+ \quad , \quad -e\phi(\boldsymbol{x}) + \beta^{-1}\ln\left(c_-(\boldsymbol{x})\Lambda^3\right) = \lambda_- \quad , \quad (14a,b)$$

that we recognize as the constant electrochemical potential conditions for the two ionic species in the solution. Thus the $\lambda_+$ and $\lambda_-$ are the ionic electrochemical potentials usually denoted by $\mu_+$ and $\mu_-$, respectively. We thereby see the physical significance of $J$: $J = F - \mu_+ N_+ - \mu_- N_-$ is the grand free energy or grand potential (denoted by $\tilde{\tilde{F}}$ in Section IV.2.2 and Appendix B). We solve (14a,b) for $c_\pm(\boldsymbol{x})$ to obtain $c_\pm(\boldsymbol{x})\Lambda^3 = \exp\left[\mp\beta e\phi(\boldsymbol{x}) + \beta\lambda_\pm\right]$. As $\phi(\boldsymbol{x})$ is short-ranged, we note that when $|\boldsymbol{x}|\to\infty$, $\phi(\boldsymbol{x})\to 0$ so that $c_+(\boldsymbol{x})$ takes the its limiting bulk-solution value $\Lambda^{-3}\exp\left(\beta\lambda_+\right)$ which we denote by $\bar{c}_+$. Similarly, $\bar{c}_- = \Lambda^{-3}\exp\left(\beta\lambda_-\right)$, so that we can eliminate $\Lambda$ and the $\lambda$'s in (14a) and (14b) to obtain the standard Boltzmann distributions for the cation and anion concentrations in a 1:1 electrolyte

$$c_{\pm}(\boldsymbol{x}) = \bar{c}_{\pm} \exp\left(\mp\beta e\phi(\boldsymbol{x})\right) \quad , \tag{15a,b}$$

in agreement with (4b), as we shall see next that $\bar{c}_+ = \bar{c}_-$. For a positive uniformly charged wall, global electroneutrality requires

$$e\left(N_- - N_+\right) = \int_S \sigma \, dS \quad , \tag{16}$$

where $N_+$ and $N_-$ are the numbers of cations and anions in solution and *S* is the $x = 0$ surface with the fixed charge. Divide (16) by $V = L^3$ and note that the right side is $\mathrm{O}(1/L)$ compared to the left, and thus negligible in the thermodynamic limit ( $N_{\pm} \to \infty$, $V \to \infty$, $N_{\pm}/V = \bar{c}_{\pm}$ fixed). Thus, in this limit we have $\bar{c}_+ = \bar{c}_- \equiv \bar{c}$.

As all ionic interactions are assumed to be Coulombic, the Poisson equation $-\varepsilon \nabla^2 \phi(\boldsymbol{x}) = \left(\rho_{\mathrm{m}}(\boldsymbol{x}) + \rho_{\mathrm{f}}(\boldsymbol{x})\right)$ is valid, so with (4a) and (15a,b) it becomes the Poisson-Boltzmann equation:

$$\varepsilon \nabla^2 \phi(\boldsymbol{x}) = e\left[\bar{c}_- \exp\left(\beta e \phi(\boldsymbol{x})\right) - \bar{c}_+ \exp\left(-\beta e \phi(\boldsymbol{x})\right)\right] - \rho_{\mathrm{f}}(\boldsymbol{x}) \quad , \tag{17}$$

in agreement with (5) in the thermodynamic limit where $\bar{c}_+ = \bar{c}_- \equiv \bar{c}$.

In applying the Gibbs variational principle above, we have assumed the laws of electrostatics (Poisson, Coulomb) and have used the variational principle to derive the "Boltzmann" part of PB theory. We can also derive the "Poisson" part from the variational principle by introducing an extended Helmholtz free energy functional $F[c_+, c_-, \phi]$ depending on three variables and minimizing with respect to all three. Thus, we use the identities[135] $f_{\mathrm{el}}(\boldsymbol{x}) = \tfrac{1}{2}\varepsilon\left(\nabla\phi(\boldsymbol{x})\right)^2 = -\tfrac{1}{2}\varepsilon\left(\nabla\phi(\boldsymbol{x})\right)^2 + \rho(\boldsymbol{x})\phi(\boldsymbol{x})$ and rewrite $F$ as

$$\begin{aligned} F[c_+, c_-, \phi] = &-\int_V \frac{1}{2}\varepsilon\left(\nabla\phi(\boldsymbol{x})\right)^2 dV + \int_V \rho(\boldsymbol{x})\phi(\boldsymbol{x})\, dV \\ &+ \beta^{-1}\int_V \left\{c_+(\boldsymbol{x})\left[\ln\left(c_+(\boldsymbol{x})\Lambda^3\right) - 1\right] + c_-(\boldsymbol{x})\left[\ln\left(c_-(\boldsymbol{x})\Lambda^3\right) - 1\right]\right\} dV \quad , \end{aligned} \tag{18}$$

where $\rho$ depends on $c_\pm$ as before. We extremize this *F* with the constraints of fixed $N_\pm$, or $J[c_+,c_-,\phi]=F[c_+,c_-,\phi]-\int_V\left(\lambda_+c_++\lambda_-c_-\right)dV$ with no constraints. Variations with respect to $c_+$ and $c_-$ yield the Boltzmann relations $c_\pm(\boldsymbol{x})=\overline{c}_\pm\exp\left(\mp\beta e\phi(\boldsymbol{x})\right)$ as before and the variation with respect to $\phi$ now yields the Poisson equation $\nabla^2\phi(\boldsymbol{x})=-\rho(\boldsymbol{x})/\varepsilon$ .

In this section we have demonstrated that the Helmholtz free energy is the relevant free energy to use in the Gibbs variational principle when the bulk concentrations require fixed numbers of ions, and fixed volume and temperature, and the boundary conditions require fixed surface charge. In later sections we show that not only a change of bulk conditions (to, for example, fixed ionic chemical potentials, or fixed pressure) necessitates the use of a Legendre transformed free energy in the variational principle, but so does a change of boundary condition from fixed surface charge to fixed surface potential.

## III. ANALYTICAL SOLUTION OF THE PB EQUATION FOR WALL GEOMETRY

We again consider a univalent 1:1 electrolyte adjacent to a uniform positively charged wall of infinite extent and simplify the PB equation (5) to the form of a one-dimensional or ordinary differential equation (with $x>0$ and a prime denoting $d/dx$)

$$\phi''(x)=(2e\overline{c}/\varepsilon)\sinh\left(\beta e\phi(x)\right) \quad . \tag{19}$$

We assume the boundary surface at $x=0$ has uniform charge density $\sigma$. On multiplying (19) throughout by $\varepsilon\phi'(x)$ we obtain

$$\frac{\varepsilon}{2}\frac{d}{dx}\phi'(x)^2=\frac{2\overline{c}}{\beta}\frac{d}{dx}\left(\cosh\left(\beta e\phi(x)\right)\right) \quad , \tag{20}$$

which is readily integrated to yield the first integral

$$\frac{\varepsilon}{2}\phi'(x)^2 - 2\bar{c}\,k_B T \cosh\left(\beta e\phi(x)\right) = const = \tfrac{1}{2}\sigma^2/\varepsilon - 2\bar{c}\,k_B T\cosh\left(\beta e\phi^0\right) \quad , \tag{21}$$

where the last form follows from taking $x = 0$ and using the boundary condition $E_x(0) = -\phi'(0) = \sigma/\varepsilon$, with $\phi^0 \equiv \phi(0)$. On noting that the electric field vanishes at infinity so that $\phi'(\infty) = 0$ and that our reference potential requires $\phi(\infty) = 0$, we see that an equivalent expression for the constant of integration is $const = -2\bar{c}\,k_B T$. Using this value for $const$ and integrating $dx = d\phi/\phi'$ from $x = 0$ to $x$, we find the inverse form of the solution of (21), $x(\phi)$, to be $x = \int_{\phi^0}^{\phi} d\phi/\phi'$, with $\phi' = -\left(4\bar{c}\,k_B T/\varepsilon\right)^{\frac{1}{2}}\left(\cosh\left(\beta e\phi\right)-1\right)^{\frac{1}{2}}$ where we have taken the negative square root here to ensure $\phi'(0) < 0$ for $\sigma > 0$. For a general $z : z'$ electrolyte this integral form would require, when generalized, elliptic integrals[101], or the equivalent direct form of the solution, $\phi(x)$, involving Jacobi elliptic functions[138]. In the case of a 1:1 electrolyte we get the relatively simple inverse form $\alpha\exp\left(-\kappa x\right) = \left(\exp\left(\beta e\phi/2\right)-1\right)/\left(\exp\left(\beta e\phi/2\right)+1\right)$, corresponding to the direct form (which can be verified to satisfy the PB equation (19) or the first integral (21))

$$\phi(x) = \frac{2k_B T}{e}\ln\left\{\frac{1+\alpha\exp(-\kappa x)}{1-\alpha\exp(-\kappa x)}\right\} \quad , \tag{22a}$$

in which $\alpha = \tanh(\beta e\phi^0/4)$ and the electrolyte screening constant $\kappa = 1/\ell_D$ is the reciprocal of the Debye screening length (1). The ionic concentrations, that follow from the Boltzmann equations $c_\pm(x) = \bar{c}\exp\left(\mp\beta e\phi(x)\right)$, are

$$c_\pm(x) = \bar{c}\left[\frac{1 \mp \alpha\exp(-\kappa x)}{1 \pm \alpha\exp(-\kappa x)}\right]^2 \quad . \tag{22b}$$

For weak fields (i.e., small $\phi^0$ or small $\sigma$), (22a) reduces to the standard exponential expression $\phi(x) = \phi^0\exp\left(-\kappa x\right)$ which also follows directly as the relevant solution of the one dimensional version of the DH equation (6) in the form (with $x>0$) $\phi''(x)=\kappa^2\phi(x)$. In this limit the ionic concentrations reduce to $c_\pm(x)=\bar{c}\left(1\mp\beta e\phi^0\exp(-\kappa x)\right)$. In the limit of large $x$ with $\phi^0$ arbitrary, the limiting values of $\phi$ and $c_\pm$ are $\phi(x)=\left(4k_\mathrm{B}T/e\right)\alpha\exp\left(-\kappa x\right)$ and $c_\pm(x)=\bar{c}\left(1\mp 4\alpha\exp\left(-\kappa x\right)\right)$.

In Fig.1 we show two plots of the PB and LPB/DH curves for $\phi(x)$. The Debye screening length is $\ell_\mathrm{D}=0.96$ nm in both panels. The surface potentials are $\phi^0=40$ mV (corresponding to a surface charge density $\sigma=0.0319$ C m$^{-2}$ obtained from (23) below) in panel A, and $\phi^0=200\,\mathrm{mV}$ (corresponding to $\sigma=0.907$ C m$^{-2}$) in panel B. Assuming $\sigma>0$ we introduce a second characteristic length $\ell_\mathrm{GC}=2\varepsilon k_\mathrm{B}T/(e\sigma)$, the Gouy-Chapman length which characterizes the strength of the surface charge. (The length $\ell_\mathrm{GC}$ will reappear in section IV.2.6.) Thus, in panel A with $\ell_\mathrm{GC}=1.12\,\mathrm{nm}$ we have $\ell_\mathrm{GC}>\ell_\mathrm{D}$ ,whereas in panel B with $\ell_\mathrm{GC}=0.039$ nm we have $\ell_\mathrm{GC}<<\ell_\mathrm{D}$. The DH curve decays over a distance $\ell_\mathrm{D}$ in both panels. In panel A, with small $\sigma$ or large $\ell_\mathrm{GC}$, the decay length of the PB curve is only slightly less than $\ell_\mathrm{D}$. By contrast, in panel B with large $\sigma$ or small $\ell_\mathrm{GC}$, the decay length of the PB curve is significantly smaller than $\ell_\mathrm{D}$ and lies between $\ell_\mathrm{GC}$ and $\ell_\mathrm{D}$. In this latter case, the stronger surface charge density pulls the counterions in more closely, causing the potential to be screened more effectively than predicted by linear DH theory, where the screening length $\ell_\mathrm{D}$ is independent of $\sigma$. In the limit where $\sigma$ and $\phi^0$ approach infinity with $\ell_\mathrm{D}$ fixed, we have $\alpha\to 1$ and $\ell_\mathrm{GC}\to 0$, and the PB decay length approaches zero together with $\ell_\mathrm{GC}$.

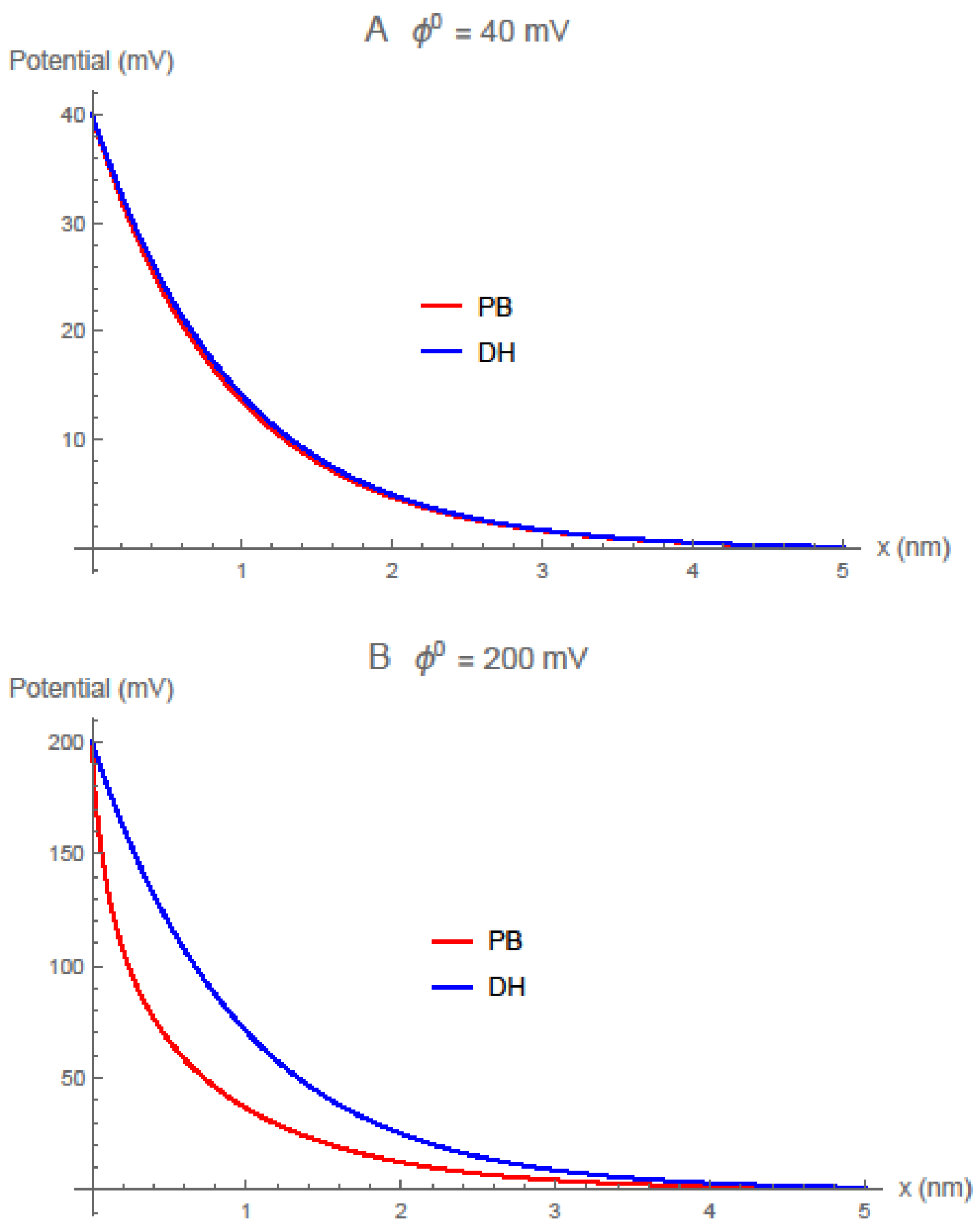

Figure 1. Plots of $\phi(x)$ vs $x$ for the PB and DH theories. In both panels we have $T = 298.15\,\mathrm{K}$, $\varepsilon_\mathrm{r} = 78.54$, $\bar{c} = 100$ mM and $\ell_\mathrm{D} = 0.96\,\mathrm{nm}$. In panel A we assume $\phi^0 = 40$ mV (corresponding to $\alpha = 0.371$) and in panel B $\phi^0 = 200$ mV (corresponding to $\alpha = 0.960$).

On equating our two equivalent expressions for *const* from the first integral we immediately obtain the Grahame[139] equation $\sigma^2 = (4\varepsilon\bar{c}\,k_\mathrm{B}T)\left[\cosh(\beta e\phi^0) - 1\right]$ which simplifies to

$$\sigma = (8\varepsilon\bar{c}\,k_\mathrm{B}T)^{\frac{1}{2}}\sinh(\beta e\phi^0/2) \quad , \tag{23}$$

an important analytic result using PB theory to predict how the surface charge density $\sigma$ is related to the surface potential $\phi^0$. An experimentally related result known as the Lippmann equation enables us to determine the surface charge density $\sigma$ of an electrified metallic droplet, such as liquid mercury, from the isothermal derivative of the surface tension $\gamma$ with respect to the surface potential $\phi^0$. Following Verwey and Overbeek[101], we note that the change in Helmholtz free energy $F$ accompanying isothermal increases in the surface area $A$ and the surface charge $q$ of an electrified liquid metal is given by $dF = \gamma dA + \phi^0 dq$. Upon performing a Legendre transform from $F$ to the new free energy $\tilde{F}$ (which also plays a key role in section IV.2.4),

$$\tilde{F} = F - \phi^0 q \quad , \tag{24}$$

we find that $d\tilde{F} = \gamma dA - q\, d\phi^0$. As this differential is exact we immediately obtain the Lippmann equation[140]

$$\left(\frac{\partial \gamma}{\partial \phi^0}\right)_{A,T} = -\left(\frac{\partial q}{\partial A}\right)_{\phi^0,T} \equiv -\sigma \ . \tag{25}$$

It predicts that the variation of the surface tension with respect to the surface potential passes through a maximum value when the surface is uncharged, i.e., when there is no electric double layer. The Lippmann equation therefore predicts that an electric double layer reduces surface tension at the metal/electrolyte interface.

One of the most studied observables for electric double layers is the differential capacitance defined by

$$C_{\mathrm{d}} \equiv \left(\partial\sigma / \partial\phi^0\right)_{A,T} = -\left(\partial^2\gamma / \partial^2\phi^0\right)_{A,T} \quad , \tag{26}$$

where the second form in (26) follows from (25). An analytical expression for $C_{\mathrm{d}}$ in PB theory therefore follows from the Grahame expression (23) as

$$C_{\mathrm{d}} = \left(2\beta e^2 \bar{c}\,\varepsilon\right)^{\frac{1}{2}} \cosh\left(\beta e \phi^0 / 2\right) \quad . \tag{27}$$

The limiting value of (27) for small $\phi^0$ is $C_{\mathrm{d}} \simeq \left(2\beta e^2 \bar{c}\,\varepsilon\right)^{\frac{1}{2}} = \varepsilon / \ell_{\mathrm{D}}$. Comparing this value with the expression $\varepsilon / \ell$ for the capacitance per unit plate area of a standard parallel plate capacitor, where $\ell$ is the plate separation, we see that the effective 'plate separation' for the small-field double layer capacitor is $\ell = \ell_{\mathrm{D}}$, a result we might have anticipated physically since $\ell_{\mathrm{D}}$ is the effective width of the double layer for small surface potential. For larger applied fields, from the parabola-like shape of $\cosh(\beta e \phi^0 / 2)$ we see that $C_{\mathrm{d}}$ is predicted to show a minimum at $\phi^0 = 0$. This parabolic behavior is in fact observed near $\phi^0 = 0$ at low electrolyte concentrations $\bar{c}$ but deviations are observed[83] at large values of $\phi^0$ and $\bar{c}$. Explaining such deviations requires extensions of PB theory and is an area of current research (see Kornyshev[109]).

## IV. FREE ENERGY AND FORCES BETWEEN TWO CHARGED PLATES

Derivation of the force between two parallel charged plates separated by an initially homogeneous electrolyte solution can be based on a hydrostatic equilibrium argument and we give this in Appendix A. These two plates can take an assigned separation $2d$ by applying external constraint forces to them. In this paper we consider forces that result from the PB primitive model alone. When the two plates are immersed in a large reservoir of electrolyte so that each plate experiences forces from both the internal and external electrolyte (our scenario B) the inter-plate force can be deduced from PB theory most simply by using the grand free energy as indicated in subsection IV.2.2 and Appendix B. If, on the other hand, the electrolyte is confined to the region between the two plates and contains fixed numbers of cations and anions - our scenario A - we can derive the relevant inter-plate forces for this two-plate problem straightforwardly from the Helmholtz free energy of the system as demonstrated next.

## 1. Scenario A. Electrolyte solution between plates of fixed charge. Helmholtz free energy and inter-plate force.

We initially consider two parallel square plates of area A separated by a fixed distance 2*d*, each plate carrying the same constant surface charge density $\sigma$, with the electrolyte solution confined to the region between the plates. The remaining containing walls are assumed uncharged. We choose the *x*-direction perpendicular to the plates, with one plate defining the $x = 0$ plane and the other the $x = 2d$ plane. We shall later consider the plate in the $x = 0$ plane to have a uniformly positive surface charge density $\sigma$ and the plate in the $x = 2d$ plane to have a uniformly negative surface charge density $-\sigma$. We show that in the latter case the inter-plate force is attractive at small plate separations but is purely repulsive in the former case when the two plates have equal charges. We shall also consider systems of varying size $2d$, with fixed $T$ and $\varepsilon$. To maintain constant solvent density as $2d$ increases we add extra solvent. For these two-plate problems our derivation in Section II of the Boltzmann equation for the case of the thermodynamic limit requires slight modification for our new system with finite dimension in the $x$ direction. For the infinite system considered previously we chose $x = \infty$ as the reference point for which we assumed $\phi(\infty) = 0$ and for our current finite system we now choose $x = d$ as the reference point for which we assume that $\phi(d) = 0$.

From (14a,b) we saw that $c_\pm(\boldsymbol{x})\Lambda^3 = \exp\left[\mp\beta e\phi(\boldsymbol{x}) + \beta\lambda_\pm\right]$ so that we have $c_\pm^{\mathrm{d}}\Lambda^3 = \exp\left(\beta\lambda_\pm\right)$ where $c_\pm^{\mathrm{d}} \equiv c_\pm(d)$. We can again eliminate $\Lambda$ and $\lambda_\pm$ from these relations to get

$$c_\pm(x) = c_\pm^{\mathrm{d}} \exp\left(\mp\beta e\phi(x)\right) \quad , \qquad \text{(28a,b)}$$

relevant to cations and anions in the two-plate scenario A problem. The PB equation (5) or (17) generalizes, from an argument analogous to that in Section II, to the two-plate form

$$\varepsilon\phi''(x) = -\rho_{\mathrm{m}}(x) = -e\left(c_+^{\mathrm{d}}\exp\left(-\beta e\phi(x)\right) - c_-^{\mathrm{d}}\exp\left(\beta e\phi(x)\right)\right) \;,\; 0 < x < 2d \;. \qquad (29)$$

We will consider the boundary conditions to be fixed surface charges, i.e., $\phi'(0^+) = -\sigma/\varepsilon$, $\phi'(2d^-) = \sigma/\varepsilon$ where $\sigma$, the charge density on both plates, is assumed to be positive. Later we indicate the required changes in the force calculation when fixed surface-potential boundary conditions are assumed.

In our scenario A the temperature $T$, the system volume $V$, and the total numbers of cations $N_+$ and anions $N_-$ are fixed and the electrolyte is restricted to the region between the two plates. All fields (e.g., $c_\pm(x),\ \rho(x),\ \phi(x),\ etc.$) depend only on $x$ provided that the plate areas are so large that edge effects can be neglected. The Helmholtz free energy $F$ can again be used to define the total free energy density $f(x)$ through

$$F/A = \int_0^{2d} f(x)\,dx \quad , \qquad (30)$$

where, as before, $f(x) = f_{\mathrm{el}}(x) + f_{\mathrm{id}}(x)$. The ideal solution term, defined by (10), can be recast, using Boltzmann relations (28a,b) and the definitions $e\left(c_+(x) - c_-(x)\right) = \rho_{\mathrm{m}}(x) = \rho(x) - \rho_{\mathrm{f}}(x)$, and then added to $f_{\mathrm{el}}(x)$ to give the total free energy density in the form

$$f(x) = -f_{\mathrm{el}}(x) + \rho_{\mathrm{f}}(x)\phi(x) + k_{\mathrm{B}}T\left[\ln\left(c_+^{\mathrm{d}}\Lambda^3\right) - 1\right]c_+(x) + k_{\mathrm{B}}T\left[\ln\left(c_-^{\mathrm{d}}\Lambda^3\right) - 1\right]c_-(x) \quad . \qquad (31)$$

We shall now show that minus the derivative of $F/A$ with respect to $2d$ gives the same expression as the force on the plate at $x = 2d$ as can be obtained by a hydrostatic equilibrium argument (see eq. (A5) of Appendix A).

The free energy density (31) is symmetric about the mid-plane $x = d$ so that $F/A = \int_0^{2d} f(x)\,dx = 2\int_0^d f(x)\,dx$ . Thus, the force per unit area on the plate at $x = 2d$, $\mathrm{f}_{2\mathrm{d}} = -\partial(F/A)/\partial(2d)$, can be obtained from $\mathrm{f}_{2\mathrm{d}} = -\partial/\partial d\int_0^d f(x)dx$ which, from (31), becomes explicitly

$$\begin{aligned}\mathrm{f}_{2\mathrm{d}} &= \frac{\partial}{\partial d}\left(\int_0^d f_{\mathrm{el}}(x)\,dx - \int_0^d \rho_{\mathrm{f}}(x)\phi(x)\,dx\right) - \frac{k_{\mathrm{B}}T}{c_+^{\mathrm{d}}}\left(\frac{\partial c_+^{\mathrm{d}}}{\partial d}\right)\int_0^d c_+(x)dx - \frac{k_{\mathrm{B}}T}{c_-^{\mathrm{d}}}\left(\frac{\partial c_-^{\mathrm{d}}}{\partial d}\right)\int_0^d c_-(x)dx \\ &= \frac{\partial}{\partial d}\int_0^d f_{\mathrm{el}}(x)\,dx - \int_0^d \rho_{\mathrm{f}}(x)\frac{\partial\phi(x)}{\partial d}dx - \frac{k_{\mathrm{B}}T}{c_+^{\mathrm{d}}}\left(\frac{\partial c_+^{\mathrm{d}}}{\partial d}\right)\int_0^d c_+(x)dx - \frac{k_{\mathrm{B}}T}{c_-^{\mathrm{d}}}\left(\frac{\partial c_-^{\mathrm{d}}}{\partial d}\right)\int_0^d c_-(x)dx \quad ,(32)\end{aligned}$$

where we have used the facts that $\rho_{\mathrm{f}}(x)$ and the integral $\int_0^d \left(c_+(x)+c_-(x)\right)dx$ are independent of *d*. The latter result, re-expressed as $\partial/\partial d\int_0^d \left(c_+(x)+c_-(x)\right)dx = 0$ together with the Boltzmann equations (28a,b) and the Leibniz rule, gives

$$-\frac{k_{\mathrm{B}}T}{c_+^{\mathrm{d}}}\left(\frac{\partial c_+^{\mathrm{d}}}{\partial d}\right)\int_0^d c_+(x)dx - \frac{k_{\mathrm{B}}T}{c_-^{\mathrm{d}}}\left(\frac{\partial c_-^{\mathrm{d}}}{\partial d}\right)\int_0^d c_-(x)dx = k_B T\left(c_+^{\mathrm{d}} + c_-^{\mathrm{d}}\right) - \int_0^d \rho_{\mathrm{m}}(x)\frac{\partial\phi(x)}{\partial d}dx \quad . \qquad (33)$$

Equation (32) therefore takes the form

$$\mathrm{f}_{2\mathrm{d}} = \frac{\partial}{\partial d}\int_0^d f_{\mathrm{el}}(x)\,dx + k_{\mathrm{B}}T\left(c_+^{\mathrm{d}} + c_-^{\mathrm{d}}\right) - \int_0^d \rho(x)\frac{\partial\phi(x)}{\partial d}dx \quad . \qquad (34)$$

From (A5) based on a hydrostatic equilibrium argument we expect $\mathrm{f}_{2\mathrm{d}}$ to be equal to the second term in (34), so that the other terms should cancel, as we now show. Using $\partial(\rho\phi)/\partial d = \rho\,\partial\phi/\partial d + \phi\,\partial\rho/\partial d$ and $\rho\phi = 2f_{\mathrm{el}}$ we find

$$\mathrm{f}_{2\mathrm{d}} = -\frac{\partial}{\partial d}\int_0^d f_{el}(x)\,dx + k_B T\left(c_+^{\mathrm{d}} + c_-^{\mathrm{d}}\right) + \int_0^d \phi(x)\frac{\partial\rho(x)}{\partial d}dx \quad . \qquad (35)$$

The identity $f_{\text{el}}(x)=\frac{\varepsilon}{2}\phi'(x)^2$ in conjunction with the Leibniz rule gives

$$\mathrm{f}_{2\mathrm{d}}=-\frac{\varepsilon}{2}\phi'(d)^2-\varepsilon\int_0^d \phi'(x)\frac{\partial\phi'(x)}{\partial d}dx+k_BT\left(c_+^{\mathrm{d}}+c_-^{\mathrm{d}}\right)+\int_0^d \phi(x)\frac{\partial\rho(x)}{\partial d}dx \quad . \tag{36}$$

Using Poisson's equation for $\rho(x)$ and integrating the last integral on the right-hand side by parts we get

$$\int_0^d \phi(x)\frac{\partial\rho(x)}{\partial d}dx=-\varepsilon\left[\phi(x)\frac{\partial\phi'(x)}{\partial d}\right]_0^d+\varepsilon\int_0^d \phi'(x)\frac{\partial\phi'(x)}{\partial d}dx \quad , \tag{37}$$

in which both boundary terms $\partial\phi'(d)/\partial d$ and $\partial\phi'(0)/\partial d$ are zero because the boundary electric fields are independent of the inter-plate spacing $2d$. Thus, (36) in conjunction with the symmetry condition $E_{\mathrm{x}}(d)\equiv-\phi'(d)=0$, simplifies to

$$\mathrm{f}_{2\mathrm{d}}=k_BT\left(c_+^{\mathrm{d}}+c_-^{\mathrm{d}}\right) \quad , \tag{38}$$

and agrees with the expression (A5) derived from a hydrostatic equilibrium argument. This inter-plate force (38) is always repulsive and depends only on the mid-plane ionic concentrations.

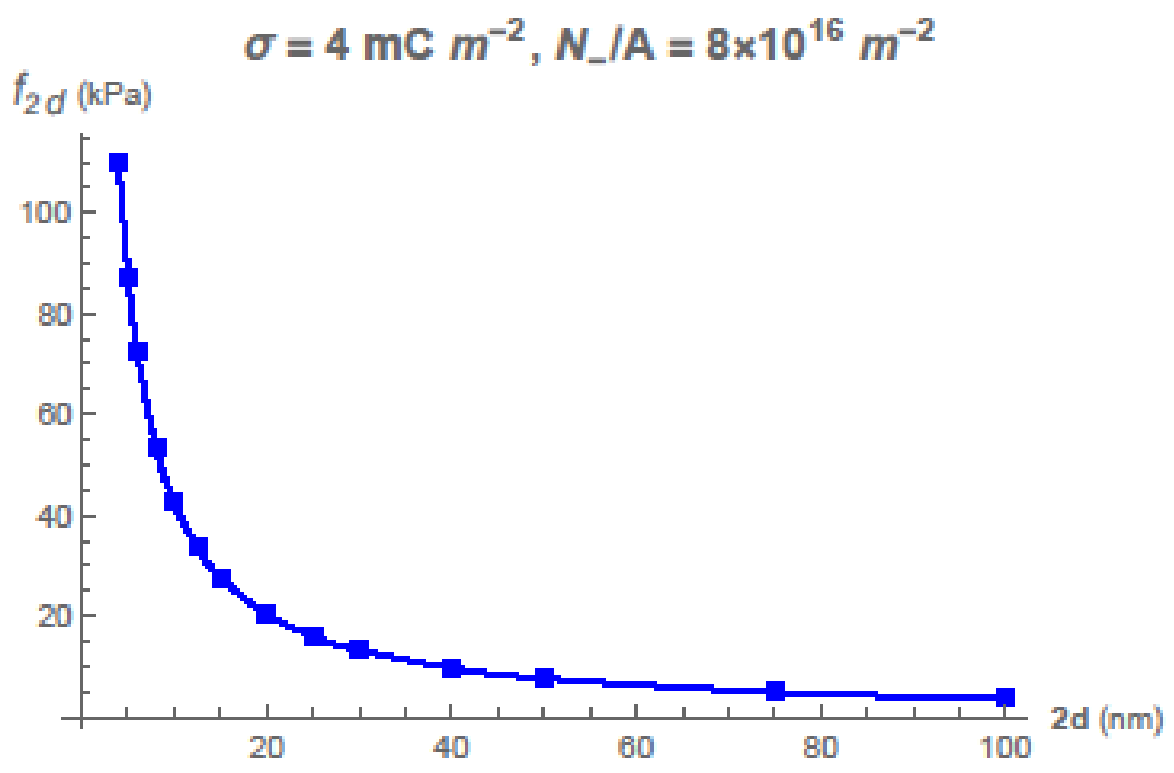

Fig.2 The inter-plate force $\mathrm{f}_{2\mathrm{d}}$ vs plate separation $2d$ for scenario A with $T=298.15\,\mathrm{K}$, $\varepsilon_{\mathrm{r}}=78.5$, $\sigma=4\,\mathrm{mC\,m^{-2}}$, and $N_-/A=8\times10^{16}\,\mathrm{m^{-2}}$. From electroneutrality $N_+/A=N_-/A-2\sigma/e=\ 3\times10^{16}\mathrm{m^{-2}}$.

In Fig.2 we show an example of the inter-plate force as a function of the plate separation $2d$. In Appendix C we give algorithms to determine the quantities $c_+^{\mathrm{d}}$ and $c_-^{\mathrm{d}}$ used to compute this force from (38). The force is purely repulsive. It can be shown to behave as $1/d$ at large separations and thus has no well-defined range.

When the two plates have surface charge densities equal in magnitude and of opposite sign, we have $\phi(d)=0$ and $c_+^{\mathrm{d}}=c_-^{\mathrm{d}}\equiv c^{\mathrm{d}}$. The sign of the inter-plate force is not now obvious, but as we shall discuss, the inter-plate force $\mathrm{f}_{2d}=2k_BT\,c^{\mathrm{d}}-(\varepsilon/2)E(d)^2$ is repulsive when the two plates are far apart and attractive when the two plates are close together.

Finally, it is worth recording the first integral of the PB equation (29) for scenario A, which can be written as

$$(\varepsilon/2)\phi'(x)^2-k_{\mathrm{B}}T\left(c_+(x)+c_-(x)\right)=const\,. \tag{39}$$

Interestingly, (39) is equivalent to the hydrostatic equilibrium relation (A4) and thus provides an alternative derivation of that relation. Again consulting Appendix A and choosing $x=0,2d$, and $d$ in (39), we find that the constant of integration in (39) takes values $\mathrm{f}_0$, $-\mathrm{f}_{2\mathrm{d}}$ and $-k_{\mathrm{B}}T\left(c_+^{\mathrm{d}}+c_-^{\mathrm{d}}\right)$, respectively, when the charges on the plates are identical. These equivalent values for $const$ also give the two-plate Grahame relation

$$\sigma^2/(2\varepsilon k_{\mathrm{B}}T)=c_+^{\mathrm{d}}\left(\exp\left(-\beta e\phi^0\right)-1\right)+c_-^{\mathrm{d}}\left(\exp\left(\beta e\phi^0\right)-1\right)\quad , \tag{40}$$

a direct generalization of the single-plate Grahame relation (23) (see also the version in the line above (23)) and to which it reduces in the limit as $d\to\infty$, $N_+\to\infty$, $N_-\to\infty$, with $N_+/(2Ad)$ and $N_-/(2Ad)$ fixed at $\bar{c}$. This is not an explicit relation between $\sigma$ and $\phi^0$ as $c_\pm^{\mathrm{d}}$ depend on $\sigma$; an explicit algorithm to determine $c_\pm^{\mathrm{d}}$ is given in Appendix C. This Grahame equation also differs from those of other scenarios, such as two plates totally immersed in the electrolyte fluid (discussed in Appendix B), two plates with opposite potentials (see Appendix C), etc.

## 2. Other Scenarios

Here we consider briefly several other interesting scenarios involving electric double layers and their interactions.

### **2.1** Scenario B

In this scenario the two parallel plates are fully immersed in the electrolyte fluid, in contrast to Scenario A discussed above where the plates form two opposing boundary walls of the fluid. This was the case first discussed in the literature, and we give a detailed description in Appendix B for two identically charged plates. For reasons discussed there we restrict ourselves, in this scenario, to thick non-conducting plates. The original application was to colloid stability; in colloid-colloid interactions, there are long-range attractive dispersion forces, but the short-range repulsive double-layer forces prevent colloid precipitation.

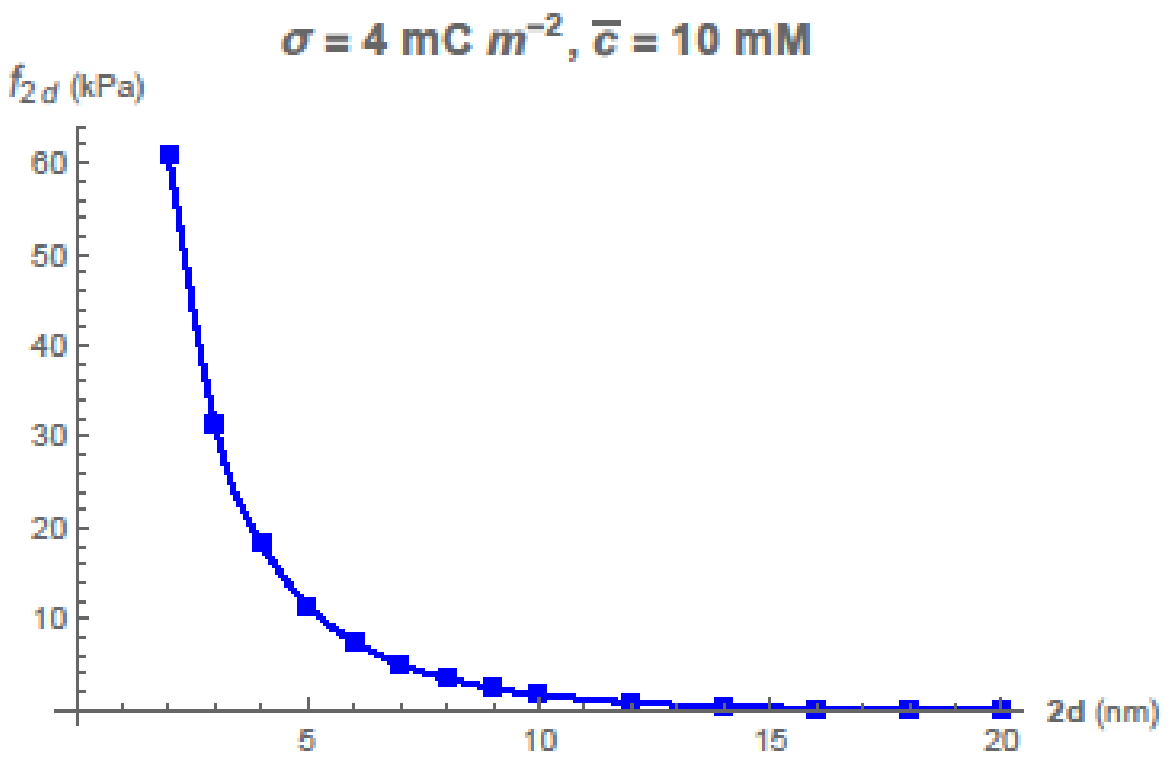

Fig.3 The inter-plate force $f_{2d}$ vs plate separation $2d$ for scenario B with $T = 298.15\,\mathrm{K}$, $\varepsilon_r = 78.5$, $\sigma = 4\,\mathrm{mC\,m^{-2}}$, and $\bar{c} = 10\,\mathrm{mM}$.

We show in Fig.3 an example of the inter-plate force $\mathrm{f}_{2\mathrm{d}}$ calculated from (B11) as a function of plate separation $2d$ . This force is seen to be purely repulsive, as in Fig.2 for scenario A. Here, however, in contrast to scenario A, the force falls off essentially as $\exp\left(-2d / \ell_{\mathrm{D}}\right)$ at large $2d$ with a well-defined range $\ell_{\mathrm{D}}$ which is about 3 nm. In Appendix C we give an algorithm for the quantity $\phi^{\mathrm{d}}$ needed to compute this force from (B11).

The case of plates with unequal surface charge density is mentioned below in subsection 2.3 and the case of plates with equal and opposite surface potentials is discussed in subsection 2.4.

## **2.2** Grand Free Energy

When the electrolyte between the plates is in equilibrium with a large reservoir of electrolyte, it is often simpler to work with the grand free energy $\tilde{\tilde{F}}$ (also called the grand potential) rather than the Helmholtz free energy $F$ that we employed in section IV.1 for scenario A. The former is related to the latter by a Legendre transformation

$$\tilde{\tilde{F}} = F - \mu_+ N_+ - \mu_- N_- \quad , \tag{41}$$

and the control variables for the transformed thermodynamic potential $\tilde{\tilde{F}}$ are the electrochemical potentials $\mu_\pm$ rather than the ion numbers $N_\pm$ relevant for $F$ . The additional control variables are *V*, *T* and $\sigma$ for both $F$ and $\tilde{\tilde{F}}$ . For fixed electrochemical potentials the ion numbers fluctuate, so that strictly speaking the $N_\pm$ in (41) are average values, but the relative fluctuations are usually negligible for a large system. In Appendix B we employ this grand free energy to analyze scenario B. From $\tilde{\tilde{F}}$ one can derive the PB equation and the inter-plate force by steps paralleling those taken above to derive these quantities from $F$ . In subsection 2.5 below we define and derive an expression for the equilibrium value of the renormalized grand free energy $\Delta\tilde{\tilde{F}}$ for a single plate.

## **2.3** Plates with Unequal Charge

In general, the left and right plates may have unequal charge, $\sigma^{\mathrm{L}} \neq \sigma^{\mathrm{R}}$. The general case is discussed by Ohshima[9] for scenario B. For scenario A with oppositely charged plates and fixed surface charge densities, $\sigma^{\mathrm{L}} = \sigma = -\sigma^{\mathrm{R}}$, the inter-plate force is given by (A5). We find the numerical calculations simpler if we employ the fixed surface potential boundary conditions considered in the next subsection; the same expression (A5) for the inter-plate force holds for fixed equal and opposite surface potentials.

## **2.4** Fixed Surface Potential Boundary Conditions

In our scenarios A and B we assumed fixed surface charge boundary conditions (fixed $\sigma$). In some problems, however, it is more convenient or more appropriate to assume fixed surface potential boundary conditions (fixed $\phi^0$). A simple example involves the two plates connected to the two terminals of a battery. The original investigators of scenario B (see Appendix B for references) employed a renormalized free energy (see following subsection) and fixed surface potential boundary conditions. The explicit boundary conditions assumed by these original investigators were fixed $\phi^0$ on both plates; with a battery supplying the external potentials the boundary conditions become fixed $\phi^0$ on one plate and fixed $-\phi^0$ on the other.

The relevant free energy $\tilde{F}$ for a fixed potential boundary condition at a single plate, together with fixed ($N_{\pm}, V, T$), is the Legendre transformation of $F$ defined by (24), or equivalently

$$\tilde{F}/A = F/A - \sigma\phi^0 \quad . \qquad (42)$$

Equation (42) has been derived by detailed chemical arguments[37,101,120], but here we give a brief alternative heuristic physical argument. Compare the Legendre transformations (41) and (42). In (41) we transform from fixed bulk ionic numbers $N_\pm$ for $F$ to fixed bulk electrochemical potentials $\mu_\pm$ for $\tilde{\tilde{F}}$, with $N_\pm$ then free to fluctuate. In (42) we transform from fixed surface charge density (fixed $\sigma$) for $F$ to fixed surface potential $\phi^0$ for $\tilde{F}$, with the surface charge $q = A\sigma$ then free to fluctuate. The role of the intensive electrostatic potential $\phi^0$ in the surface Legendre transformation (42) is analogous to the role of the intensive electrochemical potentials $\mu_\pm$ in the bulk Legendre transformation (41) of the grand free energy $\tilde{\tilde{F}}$. Thus, $\tilde{F}$ is a surface grand free energy analogous to the bulk grand free energy $\tilde{\tilde{F}}$ and in both cases the potentials ($\phi^0$ and $\mu_\pm$) are the natural control variables for grand free energies. (A more formal argument is given in the line below (24).) To derive the PB equation and the inter-plate force from a free energy, $\tilde{F}$ is the appropriate one to use if the bulk conditions are fixed numbers of ions $N_\pm$, volume and temperature, and the boundary conditions are fixed surface potentials.

If the bulk conditions are fixed electrochemical potentials $\mu_\pm$ together with fixed $(V,T)$ and the boundary conditions are fixed surface potentials, the appropriate free energy for the derivation of the PB equation and the inter-plate force is the combined Legendre transform of (41) and (42), $\tilde{\tilde{\tilde{F}}} = \tilde{\tilde{F}} - q\phi^0 = F - \mu_+ N_+ - \mu_- N_- - q\phi^0$. An analytic expression for the equilibrium value of the renormalized (see next subsection) version $\Delta\tilde{\tilde{\tilde{F}}}$ of $\tilde{\tilde{\tilde{F}}}$ for the single plate problem, in terms of its natural variable $\phi^0$ is seen from (53) to be

$$\Delta\tilde{\tilde{\tilde{F}}}/A = -8\bar{c}\,k_{\mathrm{B}}T\,\ell_{\mathrm{D}}\left(\cosh\left(\beta e\phi^0/2\right) - 1\right).$$

We turn now to the inter-plate force for a system with fixed electrochemical potentials and fixed surface-potential boundary conditions. From Section III and Appendices A and B, the expressions used to calculate the inter-plate forces in Fig. 4 are $\mathrm{f}_{2\mathrm{d}} = p(d) - (\varepsilon/2)\phi'(d)^2$ in scenario A, and $\mathrm{f}_{2\mathrm{d}} = -(\varepsilon/2)\phi'(d)^2$ in B, where $p(d) = \left(c_+^{\mathrm{d}} + c_-^{\mathrm{d}}\right)k_{\mathrm{B}}T$ is the mid-plane pressure, and as $\phi^{\mathrm{d}} = 0$ by symmetry, $c_+^{\mathrm{d}} = c_-^{\mathrm{d}} \equiv c^{\mathrm{d}}$. The reason why $p(d)$ is absent in B, which leads to a purely attractive force, is that in scenario B (unlike A) the fluid on the outside of the fully immersed plate at $x = 2d$ exerts a negative pressure force $-2\bar{c}\,k_{\mathrm{B}}T$ on the plate (expected physically, and shown formally in (B10)), and this cancels the $p(d)$ term. In scenario B we have $c_{+}^{\mathrm{d}} = c_{+}^{\mathrm{d}} \equiv \bar{c}$ as $c_{\pm}^{\mathrm{d}} / \bar{c} = \exp\left(\mp \beta e \phi^{\mathrm{d}}\right)$ with $\phi^{\mathrm{d}} = 0$ by symmetry, so that $p(d) = 2\bar{c}\, k_{\mathrm{B}} T$. Appendix C gives the algorithms used to calculate the inputs to the force calculations, i.e., $c^{\mathrm{d}}$ and $\phi'(d)$ for Fig. 4A, and $\phi'(d)$ for Fig. 4B.

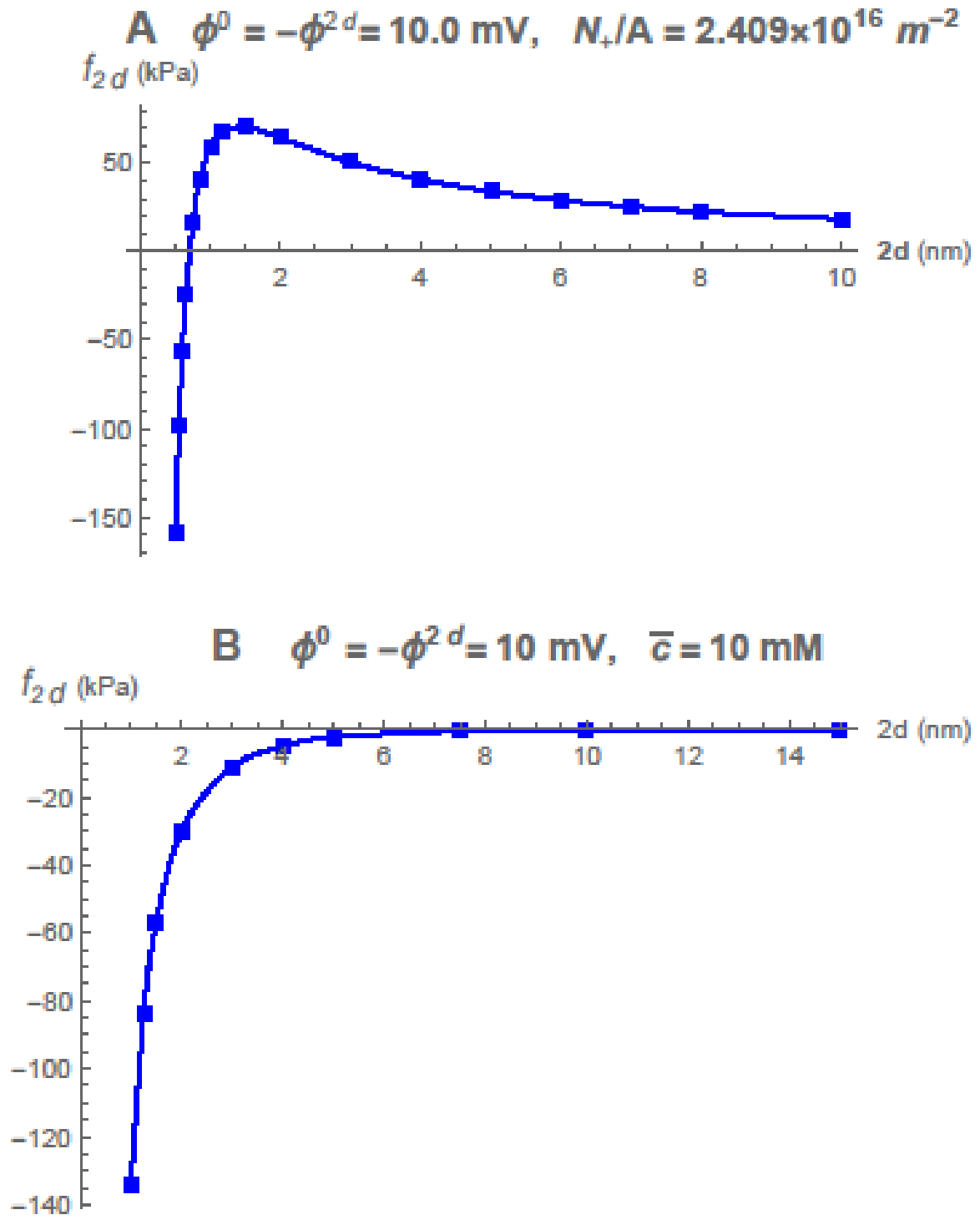

Figure 4. The inter-plate force $\mathrm{f}_{2\mathrm{d}}$ as a function of the inter-plate separation $2d$ assuming fixed surface potentials $\phi^0 = 10\,\mathrm{mV}$ on the left plate and $\phi^{2\mathrm{d}} = -10\,\mathrm{mV}$ on the right plate. Panel A refers to scenario A (with $N_+ / A = N_- / A \equiv N / A = 2.409\times10^{16}\,\mathrm{m}^{-2}$) and panel B to scenario B (with $\bar{c} = 10\ \mathrm{mM}$).

The inter-plate force is seen to have both attractive and repulsive regions in A, but is purely attractive in B. Just as pure repulsion is expected physically in Figs 2 and 3 (like charges repel), the pure attraction seen in Fig. 4B is not unexpected (unlike charges attract). The repulsive region at large $2d$ in Fig.4A deserves comment. A rough argument is that for large $2d$, $p(d)$ is of order $\left(2N_{+} / 2Ad\right) k_{\mathrm{B}} T$ and $\phi'(d)^2$ is of order $\left(2\phi^0 / 2d\right)^2$. Thus, for large values of $2d$, $\mathrm{f}_{2\mathrm{d}} = p(d) - \left(\varepsilon / 2\right) \phi'(d)^2$ is dominated by the repulsive $1/d$ pressure term.

Forces between surfaces of unlike polarity (not necessarily exactly opposite) have been studied by several groups, both experimental[131] and theoretical[128-130,132,133]. With one exception, outlined below, the scenario considered by the theoretical workers is a combination of our scenarios A and B; the mobile ions cannot access the outer surfaces of the plates (as in our scenario A) but in the region between the plates they are exchangeable with ions of a reservoir (as in our scenario B). Our results are thus not directly comparable with those of these other workers, but qualitative similarities occur. When the other workers restrict themselves to primitive model PB theory they find purely attractive forces between plates of unlike polarity, as we do in scenario B. In agreement with experiment, they find repulsion at large plate separations if they extend the theory beyond primitive model PB, or perform simulations, particularly for highly charged surfaces and multivalent mobile ions. Effects due to ion-ion Coulomb correlations, ionic finite size, and nonuniform surface charge, among others, are found to play important roles in generating the repulsive forces. The one exception is the work of McCormack et al[128] who qualitatively discuss the force for surfaces of equal and opposite potential for our scenario B in PB theory and find, as we do, a purely attractive force.

## **2.5** Renormalized Free Energy

The quantities $F$, $\tilde{F}$, $\tilde{\tilde{F}}$ and $\tilde{\tilde{\tilde{F}}}$ are absolute free energies, as defined in statistical mechanics. It is possible to shift the zero-point of these quantities. If we add a constant to any free energy, this does not affect its variation, or its derivative with respect to $d$, so the PB equation and inter-plate forces are unaffected. Many workers use such renormalized free energies, sometimes implicitly. A common renormalization is to use the excess free energy over that of the uniform system in the absence of an external electric field (no surface charges), e.g., $\Delta F = F - F^0$ for the renormalized Helmholtz free energy. With this definition $\Delta F$ can be written as $\Delta F = \int_V \left( f_{\mathrm{el}}(\boldsymbol{x}) + \Delta f_{\mathrm{id}}(\boldsymbol{x}) \right) dV$, where $f_{\mathrm{el}}(\boldsymbol{x})$ is given by any of the forms used earlier, e.g., (8).

Restricting ourselves to the equilibrium value of the free energy, and to a single plate, we first find a convenient expression for the equilibrium value of $\Delta f_{\mathrm{id}}(\boldsymbol{x})$, the renormalized ideal part of the Helmholtz free energy density. We start with the definition (10) and subtract the corresponding uniform-system value to find

$$\begin{aligned}\Delta f_{\mathrm{id}}(\boldsymbol{x}) = k_{\mathrm{B}}T\left(\ln\left(\bar{c}\Lambda^3\right)-1\right)\left(c_+(\boldsymbol{x})+c_-(\boldsymbol{x})-2\bar{c}\right)+\\ k_{\mathrm{B}}T\, c_+(\boldsymbol{x})\ln\left((c_+(\boldsymbol{x})/\bar{c}\right)+k_{\mathrm{B}}T\, c_-(\boldsymbol{x})\ln\left((c_-(\boldsymbol{x})/\bar{c}\right) \quad .\end{aligned} \tag{43}$$

We use expression (10) for $f_{\mathrm{id}}(\boldsymbol{x})$ to write the uniform-system free energy density $f^0 = 2\bar{c}k_{\mathrm{B}}T\left(\ln\left(\bar{c}\Lambda^3\right)-1\right)$, and for completeness, record the total free energy

$$F^0 = f^0 V = 2\bar{c}k_{\mathrm{B}}T\left(\ln\left(\bar{c}\Lambda^3\right)-1\right)V \quad , \tag{44}$$

of the uniform system. $F^0$ is seen to be extensive (proportional to *V*), as expected. As $\Delta F$ is a surface effect, we anticipate and find (see (50)), $\Delta F$ to be proportional to the plate area *A*.

The equilibrium Boltzmann relations $c_\pm(\boldsymbol{x})/\bar{c} = \exp\left(\mp\beta e\phi(\boldsymbol{x})\right)$, together with the definitions $ec_+(\boldsymbol{x}) - ec_-(\boldsymbol{x}) = \rho_\mathrm{m}(\boldsymbol{x}) = \rho(\boldsymbol{x}) - \rho_\mathrm{f}(\boldsymbol{x})$ and $\rho(\boldsymbol{x})\phi(\boldsymbol{x}) = 2\, f_\mathrm{el}(\boldsymbol{x})$ therefore give

$$\Delta f_\mathrm{id}(\boldsymbol{x}) = -2\, f_\mathrm{el}(\boldsymbol{x}) + \rho_\mathrm{f}(\boldsymbol{x})\phi(\boldsymbol{x}) + k_\mathrm{B}T\left(\ln\left(\bar{c}\Lambda^3\right) - 1\right)\left(c_+(\boldsymbol{x}) + c_-(\boldsymbol{x}) - 2\bar{c}\right) \quad . \qquad (45)$$

When we add $f_\mathrm{el}(\boldsymbol{x})$ to (45) we get the total renormalized free-energy density[141] $\Delta f(\boldsymbol{x})$,

$$\Delta f(\boldsymbol{x}) = -f_\mathrm{el}(\boldsymbol{x}) + \rho_\mathrm{f}(\boldsymbol{x})\phi(\boldsymbol{x}) + k_\mathrm{B}T\left(\ln\left(\bar{c}\Lambda^3\right) - 1\right)\left(c_+(\boldsymbol{x}) + c_-(\boldsymbol{x}) - 2\bar{c}\right) \quad . \qquad (46)$$

It is simpler from here on to work with the grand free energy $\tilde{\tilde{F}}$ (related to the Helmholtz free energy $F$ by (41)), and its density $\tilde{\tilde{f}}(\boldsymbol{x})$, or rather the renormalized values $\Delta\tilde{\tilde{F}}$ and $\Delta\tilde{\tilde{f}}(\boldsymbol{x}) = \Delta f(\boldsymbol{x}) - \mu_+\left(c_+(\boldsymbol{x}) - \bar{c}\right) - \mu_-\left(c_-(\boldsymbol{x}) - \bar{c}\right)$. From the large distance behaviors of (14a,b) we see that $\mu_+ = \mu_- = k_\mathrm{B}T\ln\left(\bar{c}\Lambda^3\right)$, so that we find

$$\Delta\tilde{\tilde{f}}(\boldsymbol{x}) = -f_\mathrm{el}(\boldsymbol{x}) + \rho_\mathrm{f}(\boldsymbol{x})\phi(\boldsymbol{x}) - k_\mathrm{B}T\left(c_+(\boldsymbol{x}) + c_-(\boldsymbol{x}) - 2\bar{c}\right) \quad . \qquad (47)$$

Note that the $\ln\left(\bar{c}\Lambda^3\right)$ terms have now been eliminated[141].

Using the first integral (21) of the PB equation with $const = -2\bar{c}\, k_\mathrm{B}T$ (see text below (21)) we can write $f_\mathrm{el}(\boldsymbol{x})$ as

$$f_\mathrm{el}(\boldsymbol{x}) = 2\bar{c}\, k_\mathrm{B}T\left(\cosh\left(\beta e\,\phi(\boldsymbol{x})\right) - 1\right) = k_\mathrm{B}T\left(c_+(\boldsymbol{x}) + c_-(\boldsymbol{x}) - 2\,\bar{c}\right) \ , \qquad (48)$$

with the second form in (48) following from the Boltzmann relations for $c_\pm(\boldsymbol{x})/\bar{c}$. From (47) and (48) we thus have

$$\Delta\tilde{\tilde{f}}(\boldsymbol{x}) = -4\bar{c}\, k_\mathrm{B}T\left(\cosh\left(\beta e\,\phi(\boldsymbol{x})\right) - 1\right) + \rho_\mathrm{f}(\boldsymbol{x})\phi(\boldsymbol{x}) \quad . \qquad (49)$$

The corresponding value of $\Delta\tilde{\tilde{F}} = \int_V \Delta\tilde{\tilde{f}}(\boldsymbol{x})\,dV$ is therefore

$$\Delta\tilde{\tilde{F}} / A = -4\bar{c}\, k_{\mathrm{B}}T \int_0^{\infty} \left(\cosh\left(\beta e \phi(x)\right) - 1\right) dx + \sigma\phi^0 \quad , \tag{50}$$

where we have adopted the usual geometry with direction *x* perpendicular to the surface at $x = 0$ (assumed positively charged with surface density $\sigma$ and area *A*), and we have evaluated explicitly the surface term.

We change the integration variable in (50) from $x$ to $\phi$ using $dx = d\phi / \phi'$ with $\phi'$ given in the text below (21), to get

$$\Delta\tilde{\tilde{F}} / A = -2\left(\bar{c}\, k_{\mathrm{B}}T\varepsilon\right)^{\frac{1}{2}} \int_0^{\phi^0} \left(\cosh\left(\beta e \phi\right) - 1\right)^{\frac{1}{2}} d\phi + \sigma\phi^0 \quad . \tag{51}$$

Another change of variable from $\phi$ to $y = \beta e \phi$ gives

$$\Delta\tilde{\tilde{F}} / A = -2\bar{c}\, k_{\mathrm{B}}T\left(\varepsilon k_{\mathrm{B}}T / \bar{c} e^2\right)^{\frac{1}{2}} \int_0^{y^0} \left(\cosh y - 1\right)^{\frac{1}{2}} dy + \sigma\phi^0 \quad , \tag{52}$$

where $y^0 = \beta e \phi^0$. The integral $\int_0^{y^0} \left(\cosh y - 1\right)^{\frac{1}{2}} dy$ has the value $2^{\frac{3}{2}}\left(\cosh\left(y^0 / 2\right) - 1\right)$ so that the equilibrium value of the renormalized grand free energy is

$$\Delta\tilde{\tilde{F}} / A = -8\bar{c}\, k_{\mathrm{B}}T\, \ell_{\mathrm{D}} \left(\cosh\left(\beta e \phi^0 / 2\right) - 1\right) + \sigma\phi^0 \quad . \tag{53}$$

In (53) $\Delta\tilde{\tilde{F}}$ is expressed in terms of both $\sigma$ and $\phi^0$; to express it entirely in terms of its natural variable $\sigma$ we would use the Grahame relation (23) between $\sigma$ and $\phi^0$. From (53) we can immediately write down expressions for the complete grand free energy $\tilde{\tilde{F}}$, and for other renormalized free energies such as $\Delta F / A$ and $\Delta\tilde{\tilde{\tilde{F}}} / A$. As noted in the preceding subsection, the latter quantity, expressed as a function of its natural variable $\phi^0$ is just the first term in (53), i.e., (53) without the surface term $\sigma\phi^0$. This is the free energy derived by Verwey and Overbeek[101] but they fail to make the point[141] that this free energy is $\Delta\tilde{\tilde{\tilde{F}}} / A$, not $\Delta F / A$. Corresponding derivations of various equilibrium free energies for two plates can also be carried out but the results are much more complicated[101]. Despite these complications for the free energies themselves, simple expressions for free-energy derivatives with respect to the inter-plate distance, which give the inter-plate forces, *can* be derived, as discussed with examples in section III and Appendix B.

## **2.6** Counterions Only

In this scenario there are only counterions in the solution, just enough to balance the charge on one or more plates. This case presents some new features, e.g., the electric field and counterion concentration fall off much more slowly at large distances from a plate than for a full electrolyte solution and there is no Grahame relation. Also, the PB equation can be solved analytically for a variety of both planar and non-planar geometries. The literature contains both theoretical[23,38] and experimental[38] studies.

The PB equation for this scenario has just one of the two exponential terms in (5), e.g., the one with positive exponent when the counterions are negative, corresponding to a positively charged plate. In the mathematical literature this single exponential equation is known as the Liouville equation, after the mathematician Liouville who, in 1853, posed and solved it as a purely mathematical problem in two dimensions[142-147]. It also arises in other physical problems[144,145,147,148]. One of these problems concerns the potential in a space charge of electrons near a surface from which these electrons have been thermally emitted.[148]

We restrict our brief discussion to the single plate problem with surface charge density $\sigma > 0$ on the plate at $x = 0$; Andelman[23] and Israelachvili[38] have also considered the corresponding two-plate problem (for scenario B) and the associated inter-plate force. For $x > 0$ the PB equation takes the one-dimensional Liouville form

$$\phi''(x) = (e/\varepsilon) c_-^0 \exp(\beta e \phi(x)) \qquad , \qquad (54)$$

where $c_-^0 \equiv c_-(0)$, with $c_-(x)$ the anion concentration. The boundary conditions are $\phi'(0) = -\sigma/\varepsilon$ and $\phi'(\infty) = 0$. As $\phi(x)$ behaves as $\ln(x + const)$ for large $x$ we cannot use the previous normalization $\phi(\infty)=0$. We have therefore chosen $\phi(0)\equiv\phi^0=0$ as a convenient reference potential for (54).

Multiplying (54) by $\phi'(x)$ we find its first integral to be $\frac{1}{2}\varepsilon\phi'(x)^2-k_\mathrm{B}T\,c_-(x)=const$, where $c_-(x)=c_-^0\exp(\beta e\phi(x))$. As both $\phi'(x)$ and $c_-(x)$ vanish at $x=\infty$, we see that $const=0$, so that we have

$$\tfrac{1}{2}\varepsilon\phi'(x)^2=k_\mathrm{B}T\,c_-(x) \quad . \tag{55}$$

The solution of (54) or (55) for $\phi(x)$ is easily verified to be

$$\phi(x)=-\left(2k_\mathrm{B}T/e\right)\ln\left(1+x/\ell_\mathrm{GC}\right) \quad , \tag{56}$$

where $\ell_\mathrm{GC}=2\varepsilon k_\mathrm{B}T/(e\sigma)$ is the Gouy-Chapman length introduced in section III (see discussion of Fig. 1). Note that $\phi(0)=0$ as stated above. The electric field in this case, $E_\mathrm{x}(x)=-\phi'(x)=2k_\mathrm{B}T/\left(e\left(\ell_\mathrm{GC}+x\right)\right)$, has range $\ell_\mathrm{GC}$ and falls off at large $x$ as $1/x$. This contrasts with the case of the full electrolyte solution where (see Section III) the range depends on $\ell_\mathrm{D}$ as well as $\ell_\mathrm{GC}$ (it is essentially $\ell_\mathrm{D}$ for not too large $\sigma$), and the fall-off is exponential at large $x$. Note that because $\phi^0\equiv 0$, there is no Grahame relation between surface potential and surface charge density analogous to (23) for a full electrolyte.

The corresponding anion concentration $c_-(x)=c_-^0\exp(\beta e\phi(x))$ is found, using (56), to be

$$c_-(x)=c_-^0/\left(1+x/\ell_\mathrm{GC}\right)^2 \quad , \tag{57}$$

where $c_-^0$ is found from the first integral (55) and the boundary condition $\phi'(0)=-\sigma/\varepsilon$ to be given by $c_-^0=\sigma^2/(2\varepsilon k_\mathrm{B}T)=2\varepsilon k_\mathrm{B}T/(e\ell_\mathrm{GC})^2$. The range of $c_-(x)$ is here controlled by $\ell_\mathrm{GC}$ and the fall-off at large $x$ is $1/x^2$, in contrast to the case of the full electrolyte where the range depends on both $\ell_{\rm D}$ and $\ell_{\rm GC}$ (but is approximately $\ell_{\rm D}$ for small $\sigma$) and the fall-off at large $x$ is exponential.

The equilibrium value of the Helmholtz free energy per unit plate area can be written in the form (see (31))

$$F/A = \int_0^\infty \left\{ -f_{\rm el}(x) + \rho_{\rm f}(x)\phi(x) + k_{\rm B}T\left[\ln\left(c_-^0\Lambda^3\right) - 1\right]c_-(x)\right\} dx \quad . \tag{58}$$

Since $\phi(0) = 0$, the surface term $\rho_{\rm f}(x)\phi(x)$ does not contribute to (58). The electrostatic term $f_{\rm el}(x) = \frac{1}{2}\varepsilon\phi'(x)^2$ can be expressed in terms of $c_-(x)$ using (55) and the integral $\int_0^\infty c_-(x)\,dx$ takes the value $\sigma/e$ due to global electroneutrality. An analytical expression for the free energy $F$ in terms of its natural variable $\sigma$ then follows as

$$F/A = \left(k_{\rm B}T\sigma/e\right)\left[\ln\left(c_-^0\Lambda^3\right) - 2\right], \tag{59}$$

with $c_-^0$ determined above. Note that $F \to 0$ if $\sigma \to 0$. Due to electroneutrality there are then no ions left in the solution, in contrast to the case of the full electrolyte where equal numbers of cations and anions remain in solution, and $F \to F^0$ as $\sigma \to 0$. Equation (44) for $F^0$ shows that for a full electrolyte $F$ remains non-zero as $\sigma \to 0$.

## V. CONCLUDING REMARKS

According to the fundamental variational principle of Gibbs[113,114] for thermal systems, a state of thermodynamic equilibrium minimizes the system free energy. We have shown how this principle can be applied to an electrolyte solution in contact with charged walls to obtain the nonlinear PB equation by minimizing the free energy of the system. Our discussion emphasizes that for quasi one-dimensional geometry with a uniformly charged planar wall, the nonlinear PB equation is no more complicated to solve analytically than the traditional linear DH equation considered in a few electromagnetism texts and numerous other more specialized texts. We have given the simple analytical solutions of the PB equation for both a full electrolyte such as potassium chloride in water, and for an electrolyte containing only counterions; the latter system shows some interesting novel features. Analytical expressions for the equilibrium value of the free energy are also derived. For two charged planar walls the solutions of the PB equation, and the expressions for the equilibrium values of the free energy, are much more complicated and are not derived explicitly. Despite these complications a simple expression for the derivative of the free energy with respect to the inter-wall separation, which gives the inter-wall force, *can* be derived. This expression for the force is shown to agree with that obtained from a hydrostatic equilibrium argument and is used to calculate the force for some examples as discussed below.

The nonlinear primitive PB model provides a simple theoretical description of the spatially inhomogeneous electric double layer that forms at the boundary of any electrolyte solution and the charged surface of an electrode, a biological membrane, a colloidal particle, or a charged macromolecule such as a protein or DNA. The thermodynamically preferred structure of a double layer can be predicted using the criterion that this structure minimizes the system free energy, but, as we have emphasized, care must be taken to ensure that the free energy that is minimized is consistent with which bulk and surface properties of the system have been chosen as the fixed control parameters. It is not obvious from many cases discussed in the literature whether this free energy is that of Gibbs, Helmholtz, grand or some other. Thus, for example, it is not clear from the seminal treatise of Verwey and Overbeek[101] whether their free energy of an infinite plate at constant surface potential immersed in an infinite reservoir of electrolyte refers to Helmholtz or to a grand free energy. Overbeek has since confirmed the former[126] interpretation. Our analysis, however, supports the latter interpretation. A recent attempt by Clarke and Stiles[134] to employ a free energy functional valid for fixed surface-charge boundary conditions mistakenly used a fixed surface-potential free energy functional instead. This error led them to predict a 'finite electric double layer' at the interface between a highly charged metallic plate and a 1:1 electrolyte solution, and to imply a discontinuous force per unit area at the charged interface between their 'finite electric double layer' and the bulk electrolyte. When these points are corrected, only the classical diffuse Gouy-Chapman double layer of infinite spatial extent survives.

From the perspective of the primitive model PB theory, as mentioned above forces between two parallel electrified plates separated by an electrolyte solution can be calculated from the hydrostatic equilibrium expression (A5) with a repulsive kinetic pressure component and an electrostatic component that is attractive if the plates have equal and opposite charges. Several examples are given.

In calculations of forces between two parallel charged plates, care must also be taken to distinguish between cases where the electrolyte is confined to the region between the two plates (our scenario A) and situations where the two plates are immersed in an electrolyte reservoir so that the electrolyte is in contact with both the interior and exterior surfaces of each plate (our scenario B). As shown in our explicit examples, there are quantitative (see Figs. 2 and 3) and even qualitative (see Panels A and B of Figure 4) differences between these two scenarios. Figure 4A (for scenario A) indicates inter-plate attraction between plates with equal and opposite surface potentials for small plate separations and inter-plate repulsion for large plate separations, whereas Figure 4B (for scenario B) exhibits only attractive inter-plate forces.

In conclusion, we think that PB theory deserves serious consideration for inclusion in electromagnetism courses and standard textbooks. Besides being an interesting and widely applicable aspect of nonlinear electrostatics, a subject often not mentioned, it can also serve as an excellent illustration of the power and usefulness of the fundamental Gibbs variational principle in thermodynamics and statistical mechanics.

## APPENDIX A. HYDROSTATIC EQUILIBRIUM ARGUMENT FOR INTERPLATE FORCE

In general the hydrostatic equilibrium condition is $\nabla p(\boldsymbol{x}) = \mathbf{f}_{\mathrm{B}}(\boldsymbol{x})$ where $p(\boldsymbol{x})$ is the pressure and $\mathbf{f}_{\mathrm{B}}(\boldsymbol{x})$ the body force density (per unit volume) at $\boldsymbol{x}$ due to any long-range forces acting on the fluid, e.g., gravitational or electric. For our system the body force is electric and is given by $\mathbf{f}_{\mathrm{B}}(\boldsymbol{x}) = \rho_{\mathrm{m}}(\boldsymbol{x})\boldsymbol{E}(\boldsymbol{x})$ where $\boldsymbol{E}(\boldsymbol{x})$ is the mean electric field and $\rho_{\mathrm{m}}(\boldsymbol{x})$ the mean charge density due to the mobile ions. Because of the simple wall geometry, with the charged walls perpendicular to the $x$ axis, the equilibrium condition reduces to the one-dimensional form

$$p'(x) = -\rho_{\mathrm{m}}(x)\phi'(x) \quad , \tag{A1}$$

where $p(x) = k_{\mathrm{B}}T\left(c_+(x) + c_-(x)\right)$ is the ideal or kinetic pressure due to cations and anions in the mean-field primitive model of the electrolyte. In the bulk fluid the one-dimensional Poisson equation is

$$\varepsilon\phi''(x) = -\rho_{\mathrm{m}}(x) \quad . \tag{A2}$$

Eliminating $\rho_{\mathrm{m}}(x)$ between (A1) and (A2) gives

$$p'(x) = \varepsilon\phi''(x)\phi'(x) = \frac{\varepsilon}{2}\frac{d}{dx}\phi'(x)^2 \quad . \tag{A3}$$

Integration of (A3) gives the integrated form of the hydrostatic equilibrium condition[101]

$$p(x) - \frac{\varepsilon}{2}\phi'(x)^2 = const. \tag{A4}$$

Since the left-hand side is a constant for all $x$ in the range $(0, 2d)$, we can evaluate it at $x = d$. When the two plates bear identical charges $\phi'(d) = 0$ so that $const = p(d)$, which is positive.

The force per unit area $\mathrm{f}_{2\mathrm{d}}$ on the plate at $x = 2d$ is the sum of a repulsive pressure and an attractive electrostatic contribution[153]

$$\mathrm{f}_{2\mathrm{d}} = p(2d) - \frac{\varepsilon}{2}\phi'(2d)^2 \quad . \qquad \text{(A5)}$$

This is clearly positive since (A5) is equal to $p(d)$ from (A4).

When the plates have opposite polarity, then $\phi'(d) \neq 0$, and we have $\mathrm{f}_{2\mathrm{d}} = p(d) - (\varepsilon/2)\phi'(d)^2$. In this case the sign of the net force on the plate at $x = 2d$ is not obvious; we show in section IV.2.4 and Appendix C that in scenario A the net force can be attractive or repulsive, depending on the value of $2d$, whereas in scenario B the net force is purely attractive.

As $p(d) = \left(c_+^{\mathrm{d}} + c_-^{\mathrm{d}}\right)k_{\mathrm{B}}T$, we see that determination of the inter-plate force requires $c_+^{\mathrm{d}}$ and $c_-^{\mathrm{d}}$ when the plates are equally charged, and requires $c_+^{\mathrm{d}} = c_-^{\mathrm{d}}$ and $\phi'(d)$ for oppositely charged plates. Specific algorithms for determining these quantities are discussed in Appendix C, and examples are given in sections IV.2.1 and IV.2.4.

## APPENDIX B. SCENARIO B. PLATES OF FIXED CHARGE IMMERSED IN ELECTROLYTE SOLUTION. GRAND FREE ENERGY AND INTERPLATE FORCE

Here we consider the problem, relevant to the analysis of the stability of colloidal suspensions, where each of the finite parallel plates is immersed in a large reservoir of 1:1 electrolyte, initially having a spatially uniform concentration $\overline{c}$. Rather than assume fixed numbers $N_\pm$ of mobile ions as previously, we now consider the ionic electrochemical potentials $\mu_+$ and $\mu_-$ to be fixed. The electrolyte between the two plates is assumed to be in equilibrium with the electrolyte in the reservoir and we see from the large distance behaviors of (14a,b) that $\mu_+ = \mu_- \equiv \mu = k_B T \ln\left(\bar{c}\Lambda^3\right)$. The volume $V$ and temperature $T$ are specified as before. With specified thermodynamic parameters ($T, V, \mu, \sigma$) the thermodynamic potential appropriate for use in the Gibbs variational principle and in the force calculation, is the grand free energy $\tilde{\tilde{F}} = F - \mu N_+ - \mu N_-$, where here $N_\pm$ denote the mean ionic numbers. We again neglect edge effects as the two plates are assumed to be large. To simplify this problem, we assume insulating plates with fixed inside surface charge densities $\sigma$ and double layers only on the two inside surfaces. These plates are also subject to external constraint forces in the *x*-direction that maintain the plate separation at $2d$. If the plates also carry charge on their outer surfaces, and thus have double layers there too, the inter-plate force will be unaffected. This is because the net force on the outside of a plate is unaffected by the presence of an outside double layer, as is easily seen from the hydrostatic equilibrium of any outer fluid region with one boundary on the outer surface of the plate, and the opposite planar boundary far from the plate where the concentration of each ionic species is $\bar{c}$. We also assume large plate thicknesses so that the electric field within each of the closely spaced dielectric plates tends to zero and the effects of the outer electrolyte solution on the inner one, can be neglected[9,154].

We consider circular plates, an inessential assumption but one which simplifies the description of the geometry. Consider two coaxial parallel circular plates of radius $\rho = a$ in the *yz* planes $x = 0$ and $x = 2d$. These define a cylinder of length $2d$ in the *x* direction inside a cubic container of volume $V = L^3$, with both $a$ and $L$ large, but with $a << L$ so that edge effects can be neglected and $2d << L$. We regard the total grand free energy $\tilde{\tilde{F}} = \tilde{\tilde{F}}_{\text{int}} + \tilde{\tilde{F}}_{\text{ext}}$ of the system to be the sum of an internal component $\tilde{\tilde{F}}_{\text{int}}$ corresponding to the electrolyte fluid inside the cylinder and an external component $\tilde{\tilde{F}}_{\text{ext}}$ corresponding to the fluid outside. Thus,

$$\tilde{\tilde{F}}_{\text{int}} / A = \int_0^{2d} \tilde{\tilde{f}}_{\text{int}}(x)dx \qquad , \qquad \text{(B1)}$$

and $\tilde{\tilde{F}}_{\text{ext}}$ can be taken as

$$\tilde{\tilde{F}}_{\text{ext}} / A = \int_{-L/2}^{0} \tilde{\tilde{f}}_{\text{ext}}(x)\,dx + \int_{2d}^{L/2} \tilde{\tilde{f}}_{\text{ext}}(x)\,dx \qquad , \qquad \text{(B2)}$$

where the *y* and *z* parts of the original volume integrations, leading to the purely *x*-integrals (B1) and (B2), have been restricted to the region $\rho \le a$, with area $A = \pi a^2$, where $\rho = \left(y^2 + z^2\right)^{\frac{1}{2}}$ is the cylindrical radial coordinate. When edge effects are neglected these two equations give an excellent approximation to $\tilde{\tilde{F}}$ when $\rho < a$. In the large system limit there is a negligible contribution to the inter-plate forces from external regions with $\rho > a$.

We want to calculate the net force per unit area $\mathrm{f}_{2\mathrm{d}}$ between the two plates in equilibrium with the electrolyte. The relevant PB equation again takes the form (19), but here $\phi = 0$ is chosen in the reservoir. The volume density $\tilde{\tilde{f}}_{\text{int}}(x)$ of the grand free energy per unit area $\tilde{\tilde{F}}_{\text{int}} / A$ is related to the density $f_{\text{int}}(x)$ of the Helmholtz free energy by $\tilde{\tilde{f}}_{\text{int}}(x) = f_{\text{int}}(x) - \mu c_+(x) - \mu c_-(x)$. The internal Helmholtz free energy density is $f_{\text{el}}^{\text{int}}(x) + f_{\text{id}}^{\text{int}}(x)$, where $f_{\text{el}}^{\text{int}}(x) = \frac{1}{2}\rho(x)\phi(x) = (\varepsilon / 2)\phi'(x)^2$ and $f_{\text{id}}^{\text{int}}(x)$ is defined by (10). Ionic concentrations $c_{\pm}(x)$ of the mobile ions are related to the reservoir concentration $\bar{c}$ by the Boltzmann expressions

$$c_+(x) / \bar{c} = \exp\left(-\beta e \phi(x)\right), \quad c_-(x) / \bar{c} = \exp\left(\beta e \phi(x)\right) \qquad . \qquad \text{(B3a,b)}$$

For identical charges on the two plates $\tilde{\tilde{f}}_{\text{int}}(x)$ is symmetric about the mid-plane $x = d.$ It follows that the grand free energy inside the cylinder is related to the grand free energy density by $\tilde{\tilde{F}}_{\text{int}} / A = \int_0^{2d} \tilde{\tilde{f}}_{\text{int}}(x)\,dx = 2\int_0^{d} \tilde{\tilde{f}}_{\text{int}}(x)\,dx$.

From $\tilde{\tilde{f}}_{\rm int}(x) = f_{\rm int}(x) - \mu c_+(x) - \mu c_-(x)$, the relation $\mu = k_{\rm B}T\ln\left(\bar{c}\Lambda^3\right)$, and the Boltzmann distributions (B3a,b) we find that the internal grand free energy density can be written as

$$\begin{aligned}\tilde{\tilde{f}}_{\rm int}(x) &= f_{\rm el}(x) - k_{\rm B}T\left\{c_+(x) + c_-(x) - c_+(x)\ln\left(\frac{c_+(x)}{\bar{c}}\right) - c_-(x)\ln\left(\frac{c_-(x)}{\bar{c}}\right)\right\} \\ &= \rho_{\rm f}(x)\phi(x) - f_{\rm el}(x) - k_{\rm B}T\left(c_+(x) + c_-(x)\right) \quad . \end{aligned} \qquad \text{(B4)}$$

The inter-plate force due to the internal electrolyte can now be calculated from (B4) as

$$\mathrm{f}_{2\rm d}^{\rm int} = -\left(\partial/\partial(2d)\right)_\mu \tilde{\tilde{F}}_{\rm int}/A = -\left(\partial/\partial d\right)_\mu \int_0^d \tilde{\tilde{f}}_{\rm int}(x)\,dx\,. \qquad \text{(B5)}$$

We note that $f_{\rm el}(x) = (\varepsilon/2)\phi'(x)^2$, $\rho_{\rm f}(x) = \sigma\delta(x) + \sigma\delta(x-2d)$ where the delta functions are those of Dirac, $\phi(2d) = \phi^0$ by symmetry, and that since $\mu = k_{\rm B}T\ln\left(\bar{c}\Lambda^3\right)$ the condition of fixed $\mu$ corresponds to fixed $\bar{c}$. We thus find from (B3a,b) and the Leibniz rule that (B5) becomes

$$\mathrm{f}_{2\rm d}^{\rm int} = -\sigma\frac{\partial\phi^0}{\partial d} + (\varepsilon/2)\phi'(d)^2 + \int_0^d\left[\varepsilon\,\phi'(x)\frac{\partial\phi'(x)}{\partial d} + 2k_{\rm B}T\,\bar{c}\,\frac{\partial\cosh\left(\beta e\,\phi(x)\right)}{\partial d}\right]dx + k_{\rm B}T\left(c_+^{\rm d} + c_-^{\rm d}\right). \qquad \text{(B6)}$$

From the hydrostatic equilibrium argument of Appendix A, we expect the result for $\mathrm{f}_{2\rm d}^{\rm int}$ to be just the final term in (B6), so that the other terms should cancel, which we now show. By symmetry, the mid-plane electric field $-\phi'(d) = 0$, so

$$\mathrm{f}_{2\rm d}^{\rm int} = -\sigma\frac{\partial\phi^0}{\partial d} + \int_0^d\left[\varepsilon\,\phi'(x)\frac{\partial\phi'(x)}{\partial d} + 2e\bar{c}\sinh\left(\beta e\,\phi(x)\right)\frac{\partial\phi(x)}{\partial d}\right]dx + k_{\rm B}T\left(c_+^{\rm d} + c_-^{\rm d}\right) \quad . \qquad \text{(B7)}$$

On integrating the first integral in (B7) by parts we obtain

$$\mathrm{f}_{\mathrm{2d}}^{\mathrm{int}} = -\sigma \frac{\partial \phi^0}{\partial d} + \left[ \varepsilon \phi'(x) \frac{\partial \phi(x)}{\partial d} \right]_0^d + \int_0^d \left( -\varepsilon \phi''(x) + 2e\bar{c} \sinh\left(\beta e \phi(x)\right) \right) \frac{\partial \phi(x)}{\partial d} dx + k_{\mathrm{B}} T \left( c_+^{\mathrm{d}} + c_-^{\mathrm{d}} \right). \quad \text{(B8)}$$

As, by symmetry the electric field $-\phi'(d) = 0$ and the surface charge density $\sigma = -\varepsilon \phi'(0)$, the two leading terms on the right-hand side of (B8) cancel. Also, from the PB equation (19) the integral in (B8) vanishes, so that (B8) simplifies to

$$\mathrm{f}_{\mathrm{2d}}^{\mathrm{int}} = k_{\mathrm{B}} T \left( c_+^{\mathrm{d}} + c_-^{\mathrm{d}} \right) = 2 k_{\mathrm{B}} T \bar{c} \cosh\left( \beta e \phi^{\mathrm{d}} \right). \quad \text{(B9)}$$

In calculating $\partial / \partial d \, \tilde{\tilde{F}}_{\mathrm{ext}} / A$ using (B2), note that only the second term with $x > 2d$ will contribute, and when we use the fact that $\tilde{\tilde{f}}_{\mathrm{ext}}(x) = -2 k_{\mathrm{B}} T \, \bar{c}$ is independent of *x*, we see that

$$\mathrm{f}_{\mathrm{2d}}^{\mathrm{ext}} = -\partial / \partial(2d) \left( -2 k_{\mathrm{B}} T \, \bar{c} \right) (L/2 - 2d) = -2 k_{\mathrm{B}} T \, \bar{c} \quad , \quad \text{(B10)}$$

i.e., the expected pure external pressure force pointing left.

Thus, the net force per unit area on the right plate, $\mathrm{f}_{\mathrm{2d}} = \mathrm{f}_{\mathrm{2d}}^{\mathrm{int}} + \mathrm{f}_{\mathrm{2d}}^{\mathrm{ext}}$, when the surface charges are fixed is given by

$$\mathrm{f}_{\mathrm{2d}} = k_{\mathrm{B}} T \left( c_+^{\mathrm{d}} + c_-^{\mathrm{d}} - 2\bar{c} \right) = 2 k_{\mathrm{B}} T \bar{c} \left( \cosh\left( \beta e \phi^{\mathrm{d}} \right) - 1 \right) \quad . \quad \text{(B11)}$$

Identical expressions, describing net repulsion between the plates, were established by Langmuir[156], Derjaguin and Landau[127] and Verwey and Overbeek[101] for the scenario where the plate potentials are both fixed at $\phi^0$. We briefly discuss this scenario in Section IV.2.4. As the plate separation becomes very large $\left( c_+^{\mathrm{d}} + c_-^{\mathrm{d}} \right) \to 2\bar{c}$ and we see from (B3a,b) and (B11) that $\phi^{\mathrm{d}} \to 0$ and $\mathrm{f}_{\mathrm{2d}} \to 0$.

The first integral of the PB equation (29) for scenario B has the same general form (39) as for scenario A and we find analogous values $\mathrm{f}_0^{\mathrm{int}} = -\mathrm{f}_{2\mathrm{d}}^{\mathrm{int}} = -k_{\mathrm{B}}T\left(c_+^{\mathrm{d}} + c_-^{\mathrm{d}}\right)$ for the constant of integration. From these values we find the two-plate Grahame equation

$$\sigma = \left(8\varepsilon\bar{c}\,k_{\mathrm{B}}T\right)^{\frac{1}{2}}\left(\sinh^2\left(\beta e\phi^0/2\right) - \sinh^2\left(\beta e\phi^{\mathrm{d}}/2\right)\right)^{\frac{1}{2}} \qquad \text{(B12)}$$

for scenario B, another generalization of (23) for a single plate, and to which it reduces in the limit $d \to \infty$. Unfortunately, (B12) is not a self-contained relation between $\sigma$ and $\phi^0$, as $\phi^{\mathrm{d}}$ depends on $\sigma$ (or $\phi^0$). Appendix C discusses a method for determining $\phi^{\mathrm{d}}$.

## APPENDIX C. COMPUTATION OF INTER-PLATE FORCES

In this Appendix we describe computational methods used to produce our figures[157] for inter-plate forces as a function of the plate separation $2d$. Depending on the plate polarities the inter-plate force can be assessed. Expressions describing the inter-plate forces for various conditions on the electrolyte composition and electric boundary conditions at the plates have already been derived for two identical plates and plates of opposite polarity. The inter-plate force can be expressed in terms of mid-plane ionic concentrations $c_\pm^{\mathrm{d}}$, the mid-plane electric field $E_{\mathrm{x}}(d) \equiv -\phi'(d)$, or the corresponding electric potential $\phi^{\mathrm{d}}$. As observed by Derjaguin and Landau[127] the first integral of the PB equation replaces the second-order nonlinear PB equation with appropriate boundary conditions by a simpler first-order differential equation. For the two-plate problem with identical plates the relevant Debye parameter $\kappa_{\mathrm{d}} = \left(e^2\left(c_+^{\mathrm{d}} + c_-^{\mathrm{d}}\right)/\left(\varepsilon k_{\mathrm{B}}T\right)\right)^{\frac{1}{2}}$, like the inter-plate force itself depends on the sum of the mid-plane ionic concentrations, so we can solve the first integral together with relevant physical constraints explicitly for these ionic concentrations. We consider scenarios where total ion numbers in the electrolyte solution are fixed (Figs. 2 and 4A) and where the solution ionic chemical potentials (or the reservoir electrolyte concentration) have been fixed (Figs. 3 and 4B). Whereas Figs. 2 and 3 refer to fixed surface-charge densities at the plates, Figs. 4A and 4B refer to fixed surface potentials at these plates. Apart from standard values for the fundamental physical constants we assume that the temperature $T = 298.15\,\mathrm{K}$ and that for water at $25^{\mathrm{o}}\mathrm{C}$, as solvent, the relative electric permittivity (or dielectric constant) $\varepsilon_{\mathrm{r}}$ is $78.5$.

## Figure 2. Scenario A for plates with identical surface charge densities

Input parameters include $\sigma = 4\times10^{-3}\,\mathrm{C\,m^{-2}}$ for the fixed surface charge densities and $N_{-}/A = 8\times10^{16}\,\mathrm{m^{-2}}$ for the fixed number of anions in solution per square meter of each plate. The three unknown variables of interest for a stipulated value of the plate separation $2d$ are taken to be $c_{\pm}^{\mathrm{d}}$ and the surface potential $\phi^{0}$ of either plate. As shown by (38) the mid-plane concentrations alone determine the repulsive inter-plate force. From the first integral (C1) of the relevant PB equation (29), and its boundary condition $\phi'(0) = -\sigma/\varepsilon$, we obtain the algebraic equation (40) that provides the first of three equations determining our unknown variables. On integrating $dx = d\phi/\phi'$ between $x=0$ and $x=d$ with $\phi'$ provided by the first integral of (39) with $const = -k_{\mathrm{B}}T\left(c_{+}^{\mathrm{d}} + c_{-}^{\mathrm{d}}\right)$, viz.,

$$\phi' = -\left(2k_{\mathrm{B}}T/\varepsilon\right)^{\frac{1}{2}} \left(c_{+}^{\mathrm{d}}\left(\exp(-\beta e\phi)-1\right) + c_{-}^{\mathrm{d}}\left(\exp(\beta e\phi)-1\right)\right)^{\frac{1}{2}} \quad , \tag{C1}$$

with the negative sign ensuring that the electric field at $x=0$ is positive when $\sigma > 0$, we find

$$\int_{0}^{\phi^{0}} \frac{d\phi}{\left(c_{+}^{\mathrm{d}}\left(\exp(-\beta e\phi)-1\right) + c_{-}^{\mathrm{d}}\left(\exp(\beta e\phi)-1\right)\right)^{\frac{1}{2}}} = \left(\frac{2}{\beta\varepsilon}\right)^{\frac{1}{2}} d \quad . \tag{C2}$$

This is the second of our three equations. A third independent equation can be found from electroneutrality: $N_+ = 2A\int_0^d c_+(x)\,dx = 2Ac_+^{\mathrm{d}}\int_0^d \exp\left(-\beta e\phi(x)\right)dx = N_- - 2A\sigma/e$. Upon changing the integration variable from $x$ to $\phi$, we see that this equation takes the form

$$\int_0^{\phi^0}\frac{c_+^{\mathrm{d}}\exp\left(-\beta e\phi\right)d\phi}{\left(c_+^{\mathrm{d}}\left(\exp\left(-\beta e\phi\right)-1\right)+c_-^{\mathrm{d}}\left(\exp\left(\beta e\phi\right)-1\right)\right)^{\frac{1}{2}}}=\left(\frac{2}{\beta\varepsilon}\right)^{\frac{1}{2}}\left(N_-/(2A)-\sigma/e\right)\quad . \tag{C3}$$

Using the *FindRoot* function of Wolfram's *Mathematica*[157] we have solved the algebraic equation (40) together with the two integral relations (C2) and (C3) to find $c_\pm^{\mathrm{d}}$ and $\phi^0$. The inter-plate force $\mathrm{f}_{2\mathrm{d}} = k_{\mathrm{B}}T\left(c_+^{\mathrm{d}}+c_-^{\mathrm{d}}\right)$ can then be plotted as a function of the plate separation $2d$ for the assumed values of $\sigma$ and $N_-/A$.

## Figure 3. Scenario B for plates with identical surface charge densities

As shown in (B11) the inter-plate force $\mathrm{f}_{2\mathrm{d}} = 2k_{\mathrm{B}}T\,\overline{c}\left(\cosh\left(\beta e\phi^{\mathrm{d}}\right)-1\right)$. Here, the input parameters are the surface charge density $\sigma = 4\times10^{-3}\,\mathrm{C\,m^{-2}}$ of each plate and the reservoir concentration $\overline{c} = 10\,\mathrm{mM}$. The two unknown variables of interest for a given plate separation $2d$ are taken to be the mid-plane potential $\phi^{\mathrm{d}}$ and the surface potential $\phi^0$ of each plate. They can be determined by solving the algebraic Grahame equation (B12) relevant to our scenario B together with the integral relation

$$\int_{\phi^{\mathrm{d}}}^{\phi^0}\frac{d\phi}{\left(\cosh\left(\beta e\phi\right)-\cosh\left(\beta e\phi^{\mathrm{d}}\right)\right)^{\frac{1}{2}}}=\left(\frac{4\overline{c}}{\beta\varepsilon}\right)^{\frac{1}{2}}d\quad , \tag{C4}$$

derived from $d = \int_{\phi^0}^{\phi^{\rm d}} d\phi / \phi'$, after noting that because $E_{\rm x}(d) \equiv -\phi'(d) = 0$, by symmetry, the first integral of the relevant PB equation (19) can be expressed in the form

$$\phi' = -\left(\frac{4\bar{c}}{\beta\varepsilon}\right)^{\frac{1}{2}} \left(\cosh\left(\beta e \phi\right) - \cosh\left(\beta e \phi^{\rm d}\right)\right)^{\frac{1}{2}} \quad . \tag{C5}$$

The solution to the two simultaneous equations (B12) and (C4) yields the mid-plane potential $\phi^{\rm d}$ and hence the inter-plate force for any required plate separation $2d$.

## Figure 4A. Scenario A for plates with equal and opposite surface potentials

For this case the relevant inter-plate force is given just below equation (A5) as

$$\mathrm{f}_{2\rm d} = 2k_{\rm B}T\, c^{\rm d} - \left(\varepsilon / 2\right)\phi'(d)^2 \quad . \tag{C6}$$

By symmetry, $\phi^{\rm d} = 0$, so that $c_+^{\rm d} = c_+^{\rm d} = c^{\rm d}$, and hence the ideal (kinetic) pressure becomes $p(d) = 2k_{\rm B}Tc^{\rm d}$. We fix the surface potentials of the two plates at the values $\phi^0 = 10\,\mathrm{mV}$ and $\phi^{2\rm d} = -10\,\mathrm{mV}$, and also fix the total number of cations $N_+$ in solution at $N_+ / A = 2.41\times 10^{16}\ \mathrm{m}^{-2}$. We choose our three unknown variables to be the surface charge density $\sigma$ on the left-hand plate, the mid-plane concentration $c^{\rm d}$ of cations or anions and the gradient $\phi'(d)$ of the mid-plane electrostatic potential. For a stipulated value of $d$ we initially solve two integral relations for $\sigma$ and $c^{\rm d}$, and then use the algebraic two-plate version of the Grahame equation

$$\sigma^2 / 2\varepsilon = (\varepsilon / 2)\phi'(d)^2 + 2k_{\rm B}Tc^{\rm d}\left(\cosh(\beta e\phi^0) - 1\right) \quad , \tag{C7}$$

for equal and opposite plate polarities, to obtain the second term $(\varepsilon / 2)\phi'(d)^2$ in (C6). The first of the integral relations takes the form $d = \int_{\phi^0}^{0} d\phi / \phi'$, where

$$\phi' = -\left[\left(\sigma / \varepsilon\right)^2 + (4k_{\rm B}T / \varepsilon)\, c^{\rm d}\left(\cosh(\beta e\phi - \cosh\left(\beta e\phi^0\right)\right)\right]^{\frac{1}{2}} \quad , \tag{C8}$$

so that

$$d = \int_0^{\phi^0} \frac{d\phi}{\left[(\sigma/\varepsilon)^2 + (4k_{\mathrm{B}}T/\varepsilon)\, c^{\mathrm{d}} \left(\cosh(\beta e\phi - \cosh\left(\beta e\phi^0\right)\right)\right]^{\frac{1}{2}}} \quad . \tag{C9}$$

A second integral relation in $\sigma$ and $c^{\mathrm{d}}$ comes from the relation

$$N_+ / A = \int_0^{2d} c_+(x)\, dx = \int_0^{d} \left(c_+(x) + c_-(x)\right) dx \quad , \tag{C10}$$

which can be rewritten as $N_+ / (2A) = c^{\mathrm{d}} \int_0^d \cosh\left(\beta e\,\phi(x)\right) dx$, and rewritten again, with $\phi$ as the integration variable, to give

$$N_+ / A = 2c^{\mathrm{d}} \int_0^{\phi^0} \frac{\cosh\left(\beta e\phi\right) d\phi}{\left[(\sigma/\varepsilon)^2 + (4k_{\mathrm{B}}T/\varepsilon)\, c^{\mathrm{d}} \left(\cosh(\beta e\phi - \cosh\left(\beta e\phi^0\right)\right)\right]^{\frac{1}{2}}} \quad . \tag{C11}$$

We solve (C11) and (C9) for the two unknowns $\sigma$ and $c^{\mathrm{d}}$, and then, using $(\varepsilon/2)\,\phi'(d)^2$ from (C7), we obtain the force per unit area (C6) on the right-hand plate.

## Figure 4B. Scenario B for plates with equal and opposite surface potentials

As in the previous case $\phi^{\mathrm{d}} = 0$, so that $c_+^{\mathrm{d}} = c_-^{\mathrm{d}} \equiv c^{\mathrm{d}}$. Here, however, the internal kinetic pressure $p(d) = 2c^{\mathrm{d}} k_{\mathrm{B}} T = 2\bar{c}\, k_{\mathrm{B}} T$ on the plate at $x = 2d$ due to ions between the two plates is exactly cancelled by the external pressure due to ions with $x > 2d$. Thus, in contrast to scenario A, above, the net force per unit area on the plate at $x = 2d$ takes the purely attractive form

$$\mathrm{f}_{2\mathrm{d}} = -\left(\varepsilon/2\right)\phi'(d)^2 \quad , \tag{C12}$$

directly proportional to the square of the mid-plane electric field. The first integral of the relevant PB equation (19) can be written as

$$(\varepsilon/2)\phi'^2 = \left(\varepsilon/2\right)\phi'(d)^2 + 2\bar{c}\, k_{\mathrm{B}}T\left(\cosh\left(\beta e\phi\right) - 1\right) = 2\bar{c}\, k_{\mathrm{B}}T\left(\cosh\left(\beta e\phi\right) - 1\right) - \mathrm{f}_{2\mathrm{d}} \quad . \tag{C13}$$

To relate $\phi'(d)$ to $d$ we substitute the negative square root of (C13) into $d = \int_{\phi^0}^{0} d\phi / \phi'$ to obtain

$$d = \int_0^{\phi^0} \frac{d\phi}{\left[\left(4\bar{c}\, k_B T / \varepsilon\right)\left(\cosh\left(\beta e \phi\right) - 1\right) + \phi'(d)^2\right]^{\frac{1}{2}}}$$

$$= \left(\frac{\varepsilon}{2}\right)^{\frac{1}{2}} \int_0^{\phi^0} \frac{d\phi}{\left[2\bar{c}\, k_B T\left(\cosh\left(\beta e \phi\right) - 1\right) - \mathrm{f}_{2\mathrm{d}}\right]^{\frac{1}{2}}} \quad . \qquad \text{(C14)}$$

For the fixed value $\phi^0 = 10\,\mathrm{mV}$ of the surface potential at the left-hand plate and any stipulated value of the plate separation $2d$, the integral relation (C14) can be solved numerically for the force per unit area, $\mathrm{f}_{2\mathrm{d}}$, on the right-hand plate.

## ACKNOWLEDMENT

CGG thanks NSERC (Canada) for financial support.